\documentclass[aps,pra,twocolumn,showpacs,amsmath,amssymb]{revtex4-1}
\usepackage{color}
\usepackage{graphicx}
\usepackage{dcolumn}
\usepackage{bm}
\begin{document}

\title{Non-Markovian dynamics of the electronic subsystem in a laser-driven molecule:
Characterization and connections with electronic-vibrational entanglement and electronic coherence}

\author{Mihaela Vatasescu}
\email[]{mihaela\_vatasescu@yahoo.com}
\affiliation{Institute of Space Sciences, INFLPR,
MG-23, 77125 Bucharest-Magurele, Romania}


\begin{abstract}
Non-Markovian quantum evolution of the electronic subsystem in a laser-driven molecule 
is characterized through the appearance of negative decoherence rates in the canonical form
of the electronic master equation. For a driven molecular system described in a bipartite
Hilbert space $\cal{H}$$=$$\cal{H}$$_{el}$$\bigotimes$$\cal{H}$$_{vib}$ of
dimension $2 \times N_v$, we derive the canonical form of the electronic master equation, 
deducing the canonical measures of non-Markovianity and the Bloch volume of accessible states. 
We find that one of the decoherence rates is always negative, accounting for the inherent
non-Markovian character of the electronic evolution in the vibrational environment.
Enhanced non-Markovian behavior, characterized by two negative decoherence rates,
 appears if there is a coupling between the electronic states $g$, $e$, such that the 
evolution of the electronic populations obeys $d( P_g P_e )/dt>0$. 
Non-Markovianity of the electronic evolution is analyzed in relation to
 temporal behaviors of the electronic-vibrational entanglement and electronic coherence, showing
that enhanced non-Markovian behavior accompanies entanglement increase.
Taking as an example the coupling of two electronic states by a laser pulse in the Cs$_2$ molecule,
we analyze non-Markovian dynamics under laser pulses of various strengths, 
finding that the weaker pulse stimulates the bigger amount of non-Markovianity.
Our results show that increase of the electronic-vibrational entanglement over a time interval is
correlated to the growth of the total amount of non-Markovianity calculated over the same interval
using canonical measures, and connected with the increase of the Bloch volume. After the pulse,
non-Markovian behavior is correlated to electronic coherence, such that vibrational motion 
in the electronic potentials which diminishes the nuclear overlap, implicitly increasing the linear
entropy of entanglement, brings a memory character to dynamics.
\end{abstract}


\maketitle

\section{\label{sec:intro}Introduction}

Memory effects in the dynamics of open quantum systems \cite{bookBP2002} have been extensively studied
over the past decade, through new concepts proposed to tackle quantum non-Markovianity
and examination of non-Markovian behavior in various scenarios involving open quantum systems  \cite{breuer12,rivplenio2014,breuer16,vegaalonso17}. The classical definition of a Markovian process, 
 implying a memoryless time evolution in a classic stochastic
process, cannot be simply extended to the quantum regime, where the corresponding quantum probabilities
have to be associated with measurement schemes. 
Definition of quantum Markovianity constitutes a recent research area \cite{rivplenio2014,breuer16}, 
and is still a debated subject  \cite{pollock18,chakraborty18,hallwiseman18}. 

We have to note the multiplicity of approaches to quantum non-Markovianity  \cite{rivplenio2014,breuer16,vegaalonso17,hallwiseman18}: as deviation from semigroup dynamics \cite{wolf08}, based on the backflow of information from the environment to the open system \cite{breuerlp09}, as deviation from completely positive divisibility \cite{rivas10}, based on the quantum Fisher information flow \cite{lu2010}, using entanglement-based measures \cite{rivas10,fanchini14,*debarba17} or quantum mutual-information-based measures \cite{luo12}, related to the dynamical behavior of the volume of accessible states \cite{lorenzo13}, and based on quantifiers of the negative rates in the canonical form of the time-local master equation \cite{hall2014}. Recent proposals use the spectral properties of dynamical maps \cite{chruscinski17}, and the process tensor framework \cite{pollock18,pollock18a} to characterize non-Markovian behavior. 
These alternative approaches imply different non-Markovianity concepts and propose
various measures or witnesses of quantum non-Markovianity. 
Comparative studies  \cite{addismaniscalco14,chrmaniscalco14,neto16} show them as offering different perspectives on the complex manifestation of quantum memory effects.

Non-Markovian quantum dynamics typically occurs when open quantum systems are coupled to structured or finite reservoirs, due to strong system-environment interactions, large initial system-environment correlations, or low temperature environments.  
In contrast to Markovian (memoryless) evolution of an open quantum system weakly coupled to a noisy environment, characterized by decoherence and dissipation, 
non-Markovian dynamics of an open system can lead to revivals of its characteristic quantum properties, such as quantum coherence and entanglement \cite{breuer16,huelga12,orieux15}. 
Recent developments in experimental techniques allowing control and modification of the dynamical properties of various environments through quantum reservoir engineering \cite{pciraczoller96,*myatt2000} bring forward
non-Markovian open quantum systems interacting with controllable environments \cite{liu11,tang2012,cosco18}.
These experimental advances are motivating investigations on the role of non-Markovianity as a resource for quantum information processing \cite{bylicka14,dong18} or quantum metrology \cite{escher2011,*chin12}. 
Understanding memory effects in various quantum scenarios,
such as non-Markovianity studies in driven open quantum systems \cite{schmidt15,*chen16,*poggi17,*sampaio17},
contributes to the recent attempts to design non-Markovian systems which could be useful as 
resources in quantum technologies \cite{liu11,huelga12,orieux15}.  


Molecular physics has a long tradition in treating system-bath interactions, 
including non-Markovian influences of the environment \cite{meier99,*gaspard99,*kleinek04,*welack06,*roden09,*pomyalov10}.
Non-Markovian effects operate in various molecular processes,
such as electron transfer in complex molecular systems \cite{mangaud17},
environment-assisted quantum transport \cite{reblloyd09} in molecular junctions \cite{kilgour15,*sowa17},
or excitonic energy transfer in photosynthetic complexes \cite{rebentrost09}. 
Possible applications of non-Markovianity include, for example, the use of certain molecular
 systems as quantum probes to reveal characteristic features of their environments \cite{breuer16,vegaalonso17},
or utilization of memory effects in the design of functional artificial biomaterials \cite{burghardt09}. 

Current efforts trying to exploit non-Markovianity as a resource for quantum control \cite{reich15}
rely on the understanding of memory effects as related to a backflow of information from the environment 
to the system, capable of restoring system coherence. In this sense,
recent investigations of strategies for quantum control of memory effects in molecular open-quantum systems 
 seek to protect the central system from dissipation and decoherence by increasing non-Markovian bath response \cite{puthumpal17,*puthumpal18,*mangaud18}. Non-Markovianity enhancement leading to longer decoherence times of the central system  could be exploited to increase the robustness of molecular alignment-orientation \cite{puthumpal17,*puthumpal18,*mangaud18} or to preserve coherence of molecular qubits.

Electronic coherences play an essential role in chemical and biological processes, and their function is 
currently being investigated in new domains like attochemistry or quantum biology. Recent works on 
electron dynamics in molecules explore the mechanisms influencing electronic decoherence and the role played by nuclear motion in this process, especially in the presence of strong nonadiabatic couplings \cite{vacher2017,*arnold18}. On the other hand, understanding 
quantum coherence contributions to electronic energy transport in molecular aggregates and biological systems is a major goal in quantum biology \cite{qbio2014}. Energy transport is examined using open quantum system approaches to treat 
electronic-vibrational dynamics in large molecules, in which an open "system" containing relevant  molecular electronic states is coupled to a bath of harmonic vibrational modes \cite{roden12,*rodenpre16,*rodenjcp16}.
Studies of non-Markovianity in photosynthetic complexes  have shown a significant non-Markovian information flow between electronic and phononic degrees of freedom, which could play an important role in energy transfer, as well as correlations between non-Markovian behavior
and long-lived quantum coherence \cite{ishizaki10,*aspuru11,*chen14,*chen15}.

Approaches to quantum non-Markovianity using quantum information concepts 
have been recently developed in the theory of open quantum systems, bringing new frameworks for molecular processes with memory. Non-Markovianity is recognized as a highly context-dependent concept, whose understanding should not be based solely on the evolution of the system density operator; in fact,
system-environment correlations are of direct relevance to grasp  non-Markovianity more broadly \cite{hallwiseman18}. This is also our approach here: We will characterize non-Markovianity 
of the electronic subsystem in a diatomic molecule using canonical measures, and subsequently we proceed to
understand the dynamic meaning of non-Markovian behavior by relating it
to quantum correlations in the molecular system, namely entanglement with the vibrational environment
\footnote{We emphasize that here we are referring to entanglement between the electronic system and its vibrational environment, and not to entanglement with an ancilla, proposed in Ref.~\cite{rivas10} as a non-Markovianity measure.} and electronic coherence.

We consider a diatomic molecule described in a bipartite
Hilbert space $\cal{H}$$=$$\cal{H}$$_{el}$$\bigotimes$$\cal{H}$$_{vib}$ of the electronic and vibrational degrees of freedom, driven by a laser pulse which couples the electronic states inducing transfer of population and influencing the vibrational dynamics. We shall analyze the electronic subsystem  as a driven open quantum system in the vibrational environment. 
Non-Markovianity of the electronic dynamics will be characterized
using the approach introduced by Hall {\it et al.}
in Ref.~\cite{hall2014}, which employs the canonical form of the time-local master equation describing the open system dynamics to define non-Markovianity quantifiers based on the occurrence of negative decoherence rates. We derive the canonical measures of non-Markovianity for a 2-dimensional electronic subsystem
of a laser-driven molecule, and connect non-Markovian behavior with temporal behaviors of 
electronic-vibrational entanglement (quantified using linear entropy and von Neumann entropy) and electronic coherence (measured with $l_1$ norm and Wigner-Yanase skew information). 
 
The canonical measures \cite{hall2014} provide a complete description 
of non-Markovianity in terms of canonical decoherence rates. Additionally, we shall also refer to
the Bloch volume of accessible states as a non-Markovianity witness \cite{lorenzo13,hall2014}. 
 Unlike the canonical measures, the Bloch volume is only a possible witness, and
 does not always detect non-Markovian behavior \cite{hall2014,rivplenio2014}. Nevertheless, 
examination of non-Markovianity using different measures, besides being interesting in itself, will
help to distinguish non-Markovianity regimes in the dynamical evolution,
 highlighting an "enhanced non-Markovian behavior" which is detected by both measures.

The paper is structured as follows. 
Sec.~\ref{sec:canonical} introduces the non-Markovianity approach used in this paper, based
on the occurrence of negative decoherence rates in the time-local master equation. The definitions of the canonical measures of non-Markovianity \cite{hall2014} and the Bloch volume
characterization of non-Markovianity \cite{lorenzo13,hall2014} are presented.
Sec.~\ref{sec:nonMelmol} describes our model, allowing us to characterize non-Markovian dynamics
 of the electronic subsystem of a laser-driven molecule. We derive
the canonical form of the master equation for a 2-dimensional electronic subsystem of a laser-driven molecule,
and deduce the canonical non-Markovianity measures and the Bloch volume.
Sec.~\ref{sec:nMarkov-entcoh} contains a theoretical analysis of the relations between enhancement of non-Markovianity and dynamical behaviors of the electronic-vibrational entanglement and electronic coherence.
 Sec.~\ref{sec:nMarkov-uncert} shows that 
enhanced non-Markovian behavior in the electronic evolution increases the uncertainty on the electronic energy. Sec.~\ref{sec:molecexamples} examines non-Markovian behavior of the electronic subsystem 
and its connections with electronic-vibrational entanglement and electronic coherence, taking as example the coupling of two electronic states in the Cs$_2$ molecule by laser pulses of several strengths.
The time evolutions during the pulse and after pulse are simulated numerically, being analyzed using the non-Markovianity measures, the entropies of entanglement and the measures of electronic coherence.
 Our conclusions are exposed in Sec.~\ref{sec:conclu}. The paper includes an appendix which discusses the conditions determining the increase of distinguishability between two electronic states.

\section{\label{sec:canonical} Canonical form for a local-in-time master equation and negative decoherence rates}

 The concept of quantum Markovianity implicitly used here is related to the
 concept of divisibility of dynamical maps \cite{wolfcirac08,rivplenio2014,breuer16}.
We briefly recall the notion of divisibility, which is central to the definition
of quantum (non)Markovianity in models using time-local master equations.
Considering a dynamical map $\Lambda(t,0)$ which describes the evolution $\rho(t)$ $ = \Lambda(t,0) \rho(0)$ of an open system state $\rho(t)$, $\Lambda(t,0)$ 
is a $t$-parametrized family  of completely positive and trace preserving (CPTP) maps. 
$\Lambda(t,0)$ is defined to be divisible if 
it can be written as a composition of two trace-preserving maps, 
$\Lambda(t,0)= \Lambda(t,t')\Lambda(t',0)$, for all times $t \ge t'\ge 0$, meaning that the
two-parameter family $\Lambda(t,t')$ has to exist for all $t,t'$. The positivity (P) or 
complete positivity (CP) of $\Lambda(t,t')$ lead to the notions of a P-divisible or CP-divisible 
family of dynamical maps. P divisibility and CP divisibility of a quantum process were both used 
 to define the quantum  dynamics of a process as being Markovian, and to build connections between the quantum and the classical concepts of Markovianity 
\cite{rivplenio2014,wismann15,breuer16}. Moreover, the notion of $k$-divisibility
of a dynamical map (with $1 \le k \le n$ an integer, $n$ the dimension of the open system,
$1$-divisibility corresponding to P divisibility, and $n$-divisibility corresponding to CP divisibility)
was introduced to define a "degree of non-Markovianity" of a quantum evolution  \cite{chrmaniscalco14},
as well as the notions of "weak non-Markovianity" (for processes which are only P-divisible) and 
"essential non-Markovianity" (for processes which are not even P-divisible).

A variety of theoretical and numerical methods are used to treat the dynamics of open quantum systems and to reveal the presence
of memory effects \cite{bookBP2002,breuer12,rivplenio2014,breuer16,vegaalonso17},  such as
Nakajima-Zwanzig projection operator techniques \cite{nakajima58,*zwanzig60}, the time-convolutionless (TCL) projection
operator technique \cite{shibata77,*chaturvedi79}, or stochastic wave-function techniques \cite{breuerkp99,strunz99,breuer04}. 

Quantum memory effects attached to an open system dynamics can be studied either using a
{\it non-local} master equation  with a memory kernel (obtained through the Nakajima-Zwanzig projection operator technique),
or, equivalently, using the {\it local in time} equation given by the time-convolutionless (TCL) projection
operator technique. Both approaches support an investigation of non-Markovian effects \cite{bookBP2002,chruscinski10}.
In the second approach, TCL provides a local-in-time first-order differential equation $\dot{\rho}(t) =$$\cal{L}$$(t) \rho(t)$ for the reduced density $\rho(t)$ characterizing the open system, on the condition that a certain operator inverse exists \cite{breuer12,breuer16}. For a time-local equation which does not involve a memory kernel and an integration over the
past history of the system, the non-Markovian character of the dynamics appears in the explicit time-dependence
of the generator $\cal{L}$$(t)$, which keeps the memory about the starting point \cite{breuer04,chruscinski10}.
The time-local generator $\cal{L}$$(t)$ obtained with TCL method is defined by a perturbation expansion with respect to the strength of the system-environment coupling, which does not guarantee the complete positivity of the resulting map $\Lambda(t,0)$ describing the evolution of the open system state between $0$ and $t$: 
$\rho(t)$ $ = \Lambda(t,0) \rho(0)$  \cite{bookBP2002,breuer12}.

If the requirements for preservation of the Hermiticity and the trace of $\rho(t)$ are imposed on 
the generator $\cal{L}$$(t)$ of the time-local master equation $\dot{\rho}(t) =$$\cal{L}$$(t) \rho(t)$,
one obtains a general structure of the master equation (Eq.~(\ref{canonical})),
which is a generalization of
the Gorini-Kossakowski-Sudarshan-Lindblad (GKSL) form for a memoryless
master equation \cite{breuer04,breuer12,rivplenio2014,breuer16}. 
Moreover, the diagonalization procedure leading to this GKSL-like structure provides a unique, and then canonical form of the master equation, which can be used
to characterize non-Markovianity of the time evolution ~\cite{hall2014}.

The derivation of the canonical form for a general time-local master equation $\dot{\rho}(t) =$$\cal{L}$$(t) \rho(t)$ 
 comes as a straightforward extension of the GKSL approach  \cite{gorini76,lindblad76}. We shall briefly sketch the main steps, referring to  Refs.~\cite{bookBP2002,breuer12,hall2014,rivplenio2014} for a detailed demonstration. 

Let us consider an open system described in a Hilbert space of finite dimension $d$. A complete set of 
$N:=d^2$ basis operators $\{ G_n\}_{n=0}^{N-1}$ is introduced, having the properties
\begin{eqnarray}  
G_0 = \hat{I} / \sqrt{d} ~;~ G_n = G_n^+ ~;~   \text{Tr}[G_m G_n] = \delta_{mn},
\end{eqnarray}
with $\hat{I}$ being the identity operator. $G_n$ are orthonormal traceless operators (excepting $G_0$, for which 
Tr$[G_0]=1$). A general master equation $\dot{\rho}(t) =$$\cal{L}$$(t) \rho(t)$ can be written in the following form \cite{bookBP2002,hall2014}:
\begin{eqnarray}
 \frac{d \rho}{dt}= -\frac{i}{\hbar} [ H(t), \rho(t)]\nonumber\\
 + 
\sum_{i,j=1}^{N-1} D_{ij}(t) [ G_i \rho(t) G_j - \frac{1}{2} \{G_j G_i, \rho(t) \} ],
\label{mastereq}
\end{eqnarray}
with the operator $H(t)$ being Hermitian, and $D_{ij}(t)$ being the time-dependent elements of the
 Hermitian decoherence matrix $\mathbf{D}$. The Hermiticity property of the decoherence matrix leads to the existence of 
a unique canonical form of the master equation, 
which follows using the diagonal form of $\mathbf{D}$ \cite{hall2014}:
\begin{equation}
D_{ij} (t) = \sum_{k=1}^{N-1} U_{ik} (t) \gamma_k (t) U^*_{jk} (t),
\label{diagonalD}
\end{equation}
where $\gamma_k(t)$ are the real eigenvalues of the decoherence matrix $\mathbf{D}$, and $U_{ik}(t)$ are
the elements of the unitary matrix formed by the eigenvectors of $\mathbf{D}$, such that 
$\sum_{k=1}^{N-1} U_{ik}  U^*_{jk}$$= \delta_{ij}$. Let us note that the trace of the decoherence matrix $\mathbf{D}$ equals
the sum of the {\it decoherence rates} $\gamma_k(t)$ :
\begin{equation}
\text{Tr} [ \mathbf{D} ] = \sum_k \gamma_k.
\end{equation}
If one defines the time-dependent decoherence operators $L_k(t)$ ($k=1,..,N-1$),
\begin{equation}
L_k(t) := \sum_{i=1}^{N-1} U_{ik}(t) G_i,
\label{Lkt}
\end{equation}
which form an orthonormal basis set of traceless operators
\begin{eqnarray}  
 \text{Tr}[L_j^+(t) L_k(t)] = \delta_{jk}~;~   \text{Tr}[L_k(t)] = 0,
\end{eqnarray}
Eq.~(\ref{mastereq}) can be written in the canonical form \cite{hall2014}:
\begin{eqnarray}
 \frac{d \rho}{dt}= -\frac{i}{\hbar} [ H(t), \rho]\nonumber\\ 
+ \sum_{k=1}^{N-1} \gamma_k(t) [ L_k(t) \rho L^+_k(t) - \frac{1}{2} \{ L^+_k(t)L_k(t), \rho \} ].
\label{canonical}
\end{eqnarray}
The canonical form (\ref{canonical}) is similar to the  Lindblad form of a memoryless master equation, but  the Hamiltonian $H(t)$, the decoherence operators $L_k(t)$, and the decoherence rates $\gamma_k(t)$ are time-dependent. Moreover, the decoherence operators $L_k(t)$ correspond to a set of orthogonal decoherence channels, and  the time-dependent {\it decoherence rates} $\gamma_k(t)$ obtained as eigenvalues of the decoherence matrix are uniquely determined  and can be {\it negative} \cite{hall2014}.

Formulation of necessary and sufficient conditions under which the 
 dynamics described in Eq.~(\ref{canonical}) is completely positive remains an open problem 
\cite{breuer12,breuer16}. If the rates are positive for all times, $\gamma_k(t) \ge 0$,
the dynamics is completely positive, being in Lindblad form for each fixed $t$ \cite{breuer12}.
However,
there are cases where the rates $\gamma_k(t)$ may become temporarily negative without violating complete
positivity \cite{rivplenio2014,breuer16}. 

For a master equation in the GKSL-form  (\ref{canonical}) with time-dependent coefficients, it can be shown that the corresponding dynamical map satisfies CP-divisibility if and only if  $\gamma_k(t) \ge 0$ 
\cite{rivplenio2014,breuer16}. 
The processes with a time-local master equation in the form (\ref{canonical}) and with $\gamma_k(t) \ge 0$ were also named "time-dependent Markovian" \cite{wolf08,breuerlp09} or "time-inhomogeneous Markovian" \cite{rivas10}.
To summarize, it is accepted that generalized Markovian dynamics appears for a master equation in the
quasi-GKSL-form  (\ref{canonical}) with decay rates $\gamma_k(t) \ge 0$ and a completely positive divisible dynamical map \cite{rivplenio2014,megier17}.

Non-Markovianity is related to the appearance of negative rates $\gamma_k(t) < 0$ in master equations of structure (\ref{canonical}), which leads to a violation of the divisibility property, and which was interpreted for specific systems in terms of a flow of information from the environment back to the open system \cite{breuerlp09,chrmaniscalco14}. 

It is interesting to remember the signification given to the occurrence of negative rates in models
using stochastic unraveling of time-local non-Markovian master equations \cite{breuer04,breuerkp99,strunz99}.
These models appeared as generalizations of
the stochastic wave-function method previously applied to Markovian
master equations, in order to simulate quantum master
equations with negative transition rates \cite{wisemilburn93}. In the non-Markovian quantum jumps unraveling 
\cite{piilo08},
the open system dynamics is described in terms of an ensemble of state vectors whose
non-Hermitian deterministic evolution is interrupted by random quantum jumps \cite{addismaniscalco14}.
 The time-dependent rates of the master equation are connected to the quantum jumps statistics.
The method provides an interpretation of the negative decay rates occurring in 
non-Markovian dynamics in terms of reverse quantum jumps that restore previously lost quantum superpositions \cite{piilo08}.
The negative rates reflected in reverse quantum jumps are seen as a sign of non-Markovian memory
indicating the exchange of information back and forth between the system and the reservoir \cite{piilo08}.

Hall {\it et al.} \cite{hall2014} 
have shown that for a  finite-dimensional system, the
 criterion for non-Markovianity based on the violation of CP divisibility, proposed  by 
Rivas {\it et al.} \cite{rivas10}, is
equivalent to the criterion based on the negativity of the decoherence
rates appearing in the canonical form of the master
equation.

We employ the canonical measures \cite{hall2014} to detect and quantify
non-Markovianity.  Because of their sensitivity to individual canonical decoherence rates, they 
are able to completely detect non-Markovian behavior when several decoherence channels are present.  Additionally,  the Bloch volume of accessible states is also used as a non-Markovianity witness \cite{rivplenio2014}. The two following sections expose the definitions of the canonical measures and Bloch volume, respectively.

\subsection{ \label{sec:decohrates} Negative decoherence rates and canonical measures of non-Markovianity}

Since the appearance of negative decoherence rates in the canonical form (\ref{canonical}) of the master equation
 is a feature of non-Markovianity, Hall {\it et al} \cite{hall2014} define several measures of non-Markovianity
as functions of the negative canonical decoherence rates $\gamma_k(t)$. These definitions are introduced in the following
and will be employed in our analysis.

For an individual channel $k$ with decoherence rate $\gamma_k(t)$, non-Markovianity 
can be described using the function \cite{hall2014}
\begin{eqnarray}
f_k(t) := \text{max} [0, -\gamma_k(t)] = \frac{1}{2} [ |\gamma_k(t)| -\gamma_k(t)],
\label{fkt}
\end{eqnarray}
which is 0 if the decoherence rate $\gamma_k(t)$ is positive, and $|\gamma_k(t)|$ if the decoherence rate is negative.

The canonical measure of non-Markovianity at time $t$ is defined as the sum of the
individual channels measures:
\begin{equation}
f(t) = \sum_{k}f_k(t).
\label{sumfkt}
\end{equation}
Hall {\it et al.} \cite{hall2014} have shown that their canonical measure $f(t)$ coincides, up to a multiplicative factor $2/d$ depending on the dimension $d$ of the system, with the trace-norm measure
of non-Markovianity $g(t)$ proposed by Rivas {\it et al.} \cite{rivas10}: $g(t) = 2 d^{-1} f(t)$.

One can also define a total amount of non-Markovianity in a channel $k$ over the time interval [t,t'] \cite{hall2014}  as the integral
\begin{equation}
F_k(t,t') = \int_t^{t'}f_k(s) ds,
\label{intfkt}
\end{equation}
and a total amount of non-Markovianity over the time interval [t,t'] by
\begin{equation}
F(t,t') = \sum_{k} F_k(t,t')= \int_t^{t'}f(s) ds.
\label{sumintfkt}
\end{equation}

Ref.~\cite{hall2014} also defines a non-Markov index $n(t)$ as the number of strictly negative decoherence rates:
\begin{equation}
n(t):= \# \{ k: \gamma_k(t) < 0 \}.
\label{nindex}
\end{equation}
The orthogonality of the decoherence channels allows the interpretation
of the non-Markov index $n(t)$ as the dimension of the space of non-Markovian evolution, orthogonal to the Markovian region \cite{hall2014}.

\subsection{\label{sec:Blochvol} Bloch volume characterization of non-Markovianity}

Lorenzo {\it et al.} \cite{lorenzo13} proposed a geometrical characterization of non-Markovianity based 
on the increase of the volume of states dynamically accessible to the system. 
The proposal originates in the observation that for a dynamical map corresponding to a Markovian quantum evolution the volume of physical states decreases monotonically in time, as there is no recovery of information, energy, or coherence by the system. On the contrary, a time evolution
leading to a growth in the volume of accessible states reveals physical effects associated with
non-Markovianity.

Ref.~\cite{hall2014} shows that, for a $d$-dimensional quantum system which can be represented 
by a generalized Bloch vector of dimension $d^2-1$, the Bloch volume $\mathcal{V}(t)$ at time $t$
is only sensitive to the sum of the canonical decoherence rates, $\sum_k \gamma_k(t)$, as follows:
\begin{equation}
\mathcal{V}(t) = \mathcal{V}_0 \exp  \left[ - d \int_0^t ds \sum_k \gamma_k(s) \right],
\label{Blochvol}
\end{equation}
with $\mathcal{V}_0$ being the initial volume at the time $t=0$. Consequently, the Bloch volume can 
increase at time $t$, becoming a witness of non-Markovianity, if and only if the sum of the canonical decoherence rates is negative: $\sum_k \gamma_k(t) <0$ \cite{hall2014}. Being only sensitive to the sum of the decoherence rates, there are cases when the Bloch volume cannot witness non-Markovianity 
\cite{hall2014,rivplenio2014}, as it will also appear in this paper.

\section{ \label{sec:nonMelmol} Non-Markovianity in the reduced time evolution of the electronic subsystem of a laser-driven molecule}

We will now consider the time evolution of the electronic subsystem of
a molecule driven by a laser pulse which creates entanglement between electronic and 
vibrational degrees of freedom. We treat the electronic subsystem as an open quantum system in the vibrational environment. A non-Markovian character of the electronic system dynamics is expected, since the vibrational environment is a dynamical one, being structured
by the vibrational motion in the electronic molecular potentials coupled by the laser pulse.
Therefore,  the non-Markovian effects in the electronic evolution will be
determined by the traits of the vibrational dynamics and of the driving field.
This section exposes our model, allowing us to characterize non-Markovianity of the electronic evolution
using the measures introduced in the precedent section.
We begin by describing the theoretical model of a diatomic molecule driven by a coupling 
between electronic states, such that several electronic states could be populated.
 The intramolecular dynamics of such a molecule is characterized by electronic-vibrational entanglement and electronic coherence \cite{vatasescu2013,vatasescu2015,vatasescu2016}. 
Subsequently, we will deduce the canonical form
of the master equation for a 2-dimensional electronic subsystem, building the non-Markovianity measures from the canonical decoherence rates. 

We consider a diatomic molecule described in the Born-Oppenheimer (BO) approximation \cite{bookLefebField},
neglecting the rotational degree of freedom, such that the molecular system
is described by states $|\Psi_{el,vib}(t) >$ of the Hilbert space $\cal{H}$$=$$\cal{H}$$_{el}$$\bigotimes$$\cal{H}$$_{vib}$.
 
We assume the molecule driven by the total Hamiltonian
\begin{equation}
\mathbf{\hat{H}} = \hat{H}_{mol} + \hat{W}(t),
\label{hamil}
\end{equation}
where the molecular Hamiltonian $\hat{H}_{mol}= \hat{H}_{el} + \hat{T}_R$ is the sum of 
the electronic Hamiltonian $\hat{H}_{el}$ and the
 nuclear kinetic-energy $\hat{T}_R$. $\hat{W}(t)$ describes a time-dependent coupling 
of the electronic states of the molecule \footnote{In a general manner, $\hat{W}(t)$ can be an external coupling (the case of a laser pulse), or an internal coupling (such as a radial nonadiabatic coupling between electronic states), or a combination of both.}.
The dynamics of the molecular system is obtained
from the von Neumann equation
\begin{equation}
i \hbar \frac{d \hat{\rho}_{el,vib}(t)}{dt} = [\mathbf{\hat{H}},\hat{\rho}_{el,vib}(t)],
\label{evolmol}
\end{equation}
where $\hat{\rho}_{el,vib}(t)=|\Psi_{el,vib}(t) > <\Psi_{el,vib}(t)|$ is a pure state of the bipartite system (el$\bigotimes$vib).

A detailed description of the molecular model can be found in previous papers \cite{vatasescu2013,vatasescu2015,vatasescu2016},
 where 
we have analyzed entanglement and coherence of pure states $|\Psi_{el,vib}(t) >$ created by laser pulses.
The molecular state 
$|\Psi_{el,vib}(t) >$ has the form
\begin{equation}
|\Psi_{el,vib}(t) > = \sum_{\alpha=1}^{N_{el}} |\alpha> \bigotimes |\psi_{_\alpha}(t)>,
\label{psielvibket}
\end{equation}
the summation being  over the populated electronic channels $\alpha=\overline{1,N_{el}}$. We recall that 
the molecular wave function
$\Psi_{el,vib}(\vec{r_{i}},R,t)$ depends on the  electronic coordinates $\{ \vec{r_{i}} \}$ (expressed in
the molecule-fixed coordinate system), the internuclear distance $R$,
and the time $t$. 
The electronic states $|\alpha>=\phi_{\alpha}^{el}(\vec{r_{i}};R)$ (depending  parametrically on R)
are orthonormal eigenstates of the electronic Hamiltonian $\hat{H}_{el}$ satisfying 
the clamped nuclei electronic Schr\"odinger equation $\hat{H}_{el} |\alpha>$$=V_{_\alpha}(R)|\alpha>$,
which gives the  adiabatic potential-energy surfaces $V_{_\alpha}(R)$
 as eigenvalues of $\hat{H}_{el}$ \cite{bookLefebField}. $|\psi_{_\alpha}(t)>$ designates 
the vibrational wave packet $\psi_{_\alpha}(R,t)$  corresponding to the electronic state $ |\alpha>$.

\subsection{\label{sec:elopen} The electronic subsystem as an open quantum system entangled with the vibrational environment}

We will follow the electronic subsystem dynamics in relation to dynamical behaviors of the electronic-vibrational entanglement and electronic coherence.
The reduced time evolution of the electronic subsystem is derived from the unitary dynamics
(Eq.~(\ref{evolmol}))
of the molecular system described by the molecular density operator  $\hat{\rho}_{el,vib}=|\Psi_{el,vib}(t) > <\Psi_{el,vib}(t)|$, obtained with Eq.~(\ref{psielvibket}) as
\begin{equation}
 \hat{\rho}_{el,vib}(t)= \sum_{\alpha,\beta}^{N_{el}}  |\alpha><\beta|
 \bigotimes  |\psi_{\alpha}(t) > <\psi_{\beta}(t)|.
\label{densityopN}
\end{equation}
Therefore, the reduced electronic density operator $\hat{\rho}_{el}=$Tr$_{vib}(\hat{\rho}_{el,vib})$ 
is \cite{vatasescu2015,*vatasescu2016}
\begin{equation}
 \hat{\rho}_{el}(t)= \sum_{\alpha,\beta}^{N_{el}}  | \alpha ><\beta |  <\psi_{\beta}(R,t) |\psi_{\alpha}(R,t)>.
\label{densityopelN}
\end{equation}
$\hat{\rho}_{el}(t)$ describes an electronic subsystem which is entangled with the vibrational
environment \cite{vatasescu2013}. For $N_{el}$ populated states, the linear entropy 
$L(t)=1-\text{Tr}_{el}(\hat{\rho}^2_{el}(t))$ of the electronic-vibrational 
entanglement has the expression \cite{vatasescu2015,*vatasescu2016}:
\begin{equation}
   L(t) = 2 \sum_{\alpha,\beta, \alpha \ne \beta}^{N_{el}} [ P_{_\alpha}(t) P_{_\beta}(t) -
   |<\psi_{\alpha}(R,t)|\psi_{\beta}(R,t)>|^2 ].
   \label{linentrgen}
\end{equation}
In Eq.~(\ref{linentrgen}),  $P_{_\alpha}(t)$$=$$<\psi_{\alpha}(R,t)|\psi_{\alpha}(R,t)>$ is the population of the  electronic state $|\alpha>$, and the total population obeys the normalization condition
$\sum_{\alpha=1}^{N_{el}}P_{_\alpha}(t)=1$. The other term appearing in Eq.~(\ref{linentrgen}) 
involves the off-diagonal elements 
$<\alpha|\hat{\rho}_{el}(t)|\beta>$$=<\psi_{\beta}(R,t)|\psi_{\alpha}(R,t)>$, which are giving 
the coherence of the reduced electronic state $\hat{\rho}_{el}(t)$. Using the $l_1$ norm definition of coherence \cite{baumgratz14}, 
one obtains as measure of the electronic coherence:
\begin{equation}
C_{l_1}(\hat{\rho}_{el})=\sum_{\alpha,\beta, \alpha \ne \beta}^{N_{el}} |<\psi_{\alpha}(R,t)|\psi_{\beta}(R,t)>|.
\label{l1normgen}
\end{equation}

In the following we suppose an electronic subsystem of dimension dim$(\cal{H}$$_{el})=2$,
and we derive the canonical form of the master
 equation which describes its evolution.

\subsection{ \label{sec:mastereq2} The master equation for a two-dimensional driven electronic subsystem}

We consider a diatomic molecule in which two electronic states $|g>,|e>$ are coupled by 
a laser pulse, such that a pure molecular state $|\Psi_{el,vib}(t) >$ is created:
\begin{equation}
|\Psi_{el,vib}(t) > = |g> \bigotimes |\psi_{g}(R,t)>
+ |e> \bigotimes |\psi_{e}(R,t)>.
\label{2pure-elvib}
\end{equation} 
The quantum dynamics of the molecular system driven by the Hamiltonian (\ref{hamil}) is given
by the time-dependent Schr\"odinger equation:
\begin{equation}
i \hbar \frac{\partial}{\partial t} |\Psi_{el,vib}(t) > = [\hat{H}_{mol} + \hat{W}(t)] |\Psi_{el,vib}(t) >.
\label{eqSelvib}
\end{equation} 
Projecting Eq.~(\ref{eqSelvib}) on the electronic states $|g>, |e>$, and taking into account the BO approximation
(i.e. $<\alpha|\hat{H}_{mol}|\alpha >$$=\hat{T}_R + V_{_\alpha}(R)$ and  $<\alpha|\hat{H}_{mol}|\beta >=0$),
as well as the off-diagonal nature of the coupling (i.e. $<\alpha|\hat{W}(t)|\alpha >=0$), where
$|\alpha >,|\beta>$ generically designate the electronic adiabatic states, one obtains
\begin{eqnarray}
\label{eqS2states}
&&i\hbar\frac{\partial}{\partial t}\left(\begin{array}{c}
 \psi_{g}(R,t)\\
\psi_{e}(R,t)
 \end{array}\right)=\\
&&
\left(\begin{array}{lc}
 {\hat T_R} + V_{g}(R)  & W(R,t) \\
 W^*(R,t)  & {\hat T_R} + V_e(R)
 \end{array} \right)
 \left( \begin{array}{c}
 \psi_{g}(R,t)\\
\psi_{e}(R,t)
 \end{array} \right). \nonumber
\end{eqnarray}
Eq.~(\ref{eqS2states}) describes the vibrational dynamics 
 of the  wave packets $\psi_{g,e}(R,t)$ moving in the electronic 
potentials $V_{g}(R)$ and $V_{e}(R)$, which are coupled by $W(R,t)$$=<g|\hat{W}(t)|e>$, depending on the internuclear distance 
$R$ and on the time $t$. As we have mentioned, Eq.~(\ref{eqS2states}) can be used to describe evolution 
in the case of an external driving field (we will consider a laser pulse \footnote{The theoretical model treating the 
relative motion of the nuclei in the ground and excited electronic channels coupled by a 
laser pulse usually contains supplementary assumptions, involving the rotating wave approximation
 and dressed electronic potentials \cite{vatasescu01,vatasescu09,vatasescu12}.}), as well as for an internal coupling (i.e. a radial nonadiabatic coupling between electronic states).

The matrix of the reduced electronic density $\hat{\rho}_{el}(t)$ in the electronic 
basis $\{|g>,|e>\}$ can be deduced from Eq.~(\ref{densityopelN}) as
\begin{eqnarray}
( \hat{\rho}_{el} (t) )_{ \{ g,e \} } 
=
\left(\begin{array}{cc}
   P_g(t) &  <\psi_{e}(t)|\psi_{g}(t)>\\
  <\psi_{g}(t)|\psi_{e}(t)> & P_e (t)
 \end{array} \right),\nonumber \\
\label{matdensel2}
\end{eqnarray}
where $P_{g,e}(t)$$=<\psi_{g,e}(R,t)|\psi_{g,e}(R,t)>$
are the populations of
the two electronic states $g,e$, with the normalization condition $P_g(t)+P_e(t)=1$.
From Eq.~(\ref{matdensel2}) we obtain the master equation for $\hat{\rho}_{el}(t)$,
 having the following local-in-time form:
\begin{eqnarray}
 i \hbar \frac{d \hat{\rho}_{el}}{dt}= A(t) |g><g| - A(t) |e><e|\nonumber \\
 + B(t) |g><e| - B^*(t) |e><g|.
\label{mastereldens}
\end{eqnarray}
$A(t)$ and $B(t)$ are the complex time-dependent functions
\begin{eqnarray}
 A(t) = i \hbar \frac{d P_g}{dt} = -i \hbar \frac{d P_e}{dt}, \label{At} \\
 B(t) = i \hbar \frac{d<\psi_{e}|\psi_{g}>}{dt}, \label{Bt}
\end{eqnarray}
which are determined by the time evolution of the vibrational wave packets
 $|\psi_{g}(R,t)>$ and $|\psi_{e}(R,t)>$, directed by Eq.~(\ref{eqS2states}).

The next section shows the derivation of the canonical form for Eq.~(\ref{mastereldens}).

\subsection{\label{sec:canon2el} Canonical form of the master equation for the two-dimensional electronic
 subsystem of a molecule driven by a laser pulse}

We shall derive here the canonical form of the master equation for the 2-dimensional
electronic subsystem $\hat{\rho}_{el}(t)$. The master equation (\ref{mastereldens})
wil be used to deduce both (\ref{mastereq}) and (\ref{canonical}) forms,
 in order to obtain the decoherence matrix $\mathbf{D}$ and the decoherence rates $\gamma_k(t)$. 

As dim($\cal{H}$$_{el}$)$=2$,  the orthornormal basis $\{ G_i\}_{i=0}^{3}$ can be chosen as 
$\{ \hat{I} / \sqrt{2} , \sigma_i / \sqrt{2} \}$, with $\{ \sigma_i \}_{i=1,2,3}$ being
the Pauli operators: $\sigma_1$$ = |e><g| +  |g><e|$, $\sigma_2$$ = -i|e><g| + i |g><e|$, and
$\sigma_3$$ = |e><e| -  |g><g|$.
We also use the operators $\sigma_+ = |e><g|$$ = 1/2 (\sigma_1 + i \sigma_2)$ and 
$\sigma_- = |g><e|$$ = 1/2 (\sigma_1 - i \sigma_2)$, leading to
$|g><g| = \sigma_- \sigma_+$ and $|e><e| = \sigma_+ \sigma_-$. 
As a first step, Eq.~(\ref{mastereldens}) can be written as
\begin{eqnarray}
 i \hbar \frac{d \hat{\rho}_{el}}{dt}= \frac{A(t)}{P_e} \sigma_- \hat{\rho}_{el} \sigma_+ 
 - \frac{A(t)}{P_g} \sigma_+ \hat{\rho}_{el} \sigma_-\nonumber \\ 
+ \frac{B(t)}{<\psi_{g}|\psi_{e}>} \sigma_-\hat{\rho}_{el} \sigma_-
 - \frac{B^*(t)}{<\psi_{e}|\psi_{g}>} \sigma_+ \hat{\rho}_{el} \sigma_+,
\label{mastereq1}
\end{eqnarray}
giving
\begin{equation}
 \frac{d \hat{\rho}_{el}}{dt}=  
\sum_{i,j=1,2}d_{ij}(t) \sigma_i \hat{\rho}_{el} \sigma_j.
\label{mastereq2}
\end{equation}
The form (\ref{mastereq2}) can be completed in order to sort out an equation having the structure of Eq.~(\ref{mastereq}) which  provides the decoherence matrix. 
We then obtain 
\begin{eqnarray}
 \frac{d \hat{\rho}_{el}}{dt}= -\frac{i}{\hbar} [ H(t), \hat{\rho}_{el}(t)]\nonumber \\
 + 
\sum_{i,j=1}^{3} d_{ij}(t) [ \sigma_i \hat{\rho}_{el}(t) \sigma_j -
 \frac{1}{2} \{\sigma_j \sigma_i, \hat{\rho}_{el}(t) \} ].
\label{mastereqeldens}
\end{eqnarray}
In Eq.~(\ref{mastereqeldens}),
the Hermitian operator  $H(t)$  has the following matrix in the electronic basis $\{|g>,|e>\}$:
\begin{eqnarray}
( H(t))_{ \{ g,e \} } 
=\nonumber \\
- \left(\begin{array}{cc}
   P_g \frac{Re(<\psi_{g}| W|\psi_{e}>)}{|<\psi_{g}|\psi_{e}>|^2}    &  \frac{<\psi_{g}| W |\psi_{e}> }{<\psi_{g}|\psi_{e}>}
\\ \\
  \frac{<\psi_{e}| W^* |\psi_{g}> }{<\psi_{e}|\psi_{g}>} & P_e \frac{Re(<\psi_{g}| W|\psi_{e}>)}{|<\psi_{g}|\psi_{e}>|^2}
 \end{array} \right),
\label{Hmatrix}
\end{eqnarray}
and the matrix $d_{ij}(t)$ has the form
\begin{eqnarray}
(d_{ij}(t)) =
\left(\begin{array}{ccc}
 d_{11}(t)  &
 d_{12}(t) &
0 \\
d_{21}(t) &
d_{22}(t) &
0  \\
 0  &
0  &
 d_{33}(t)
 \end{array} \right).
\label{matrixd}
\end{eqnarray}
The elements $D_{ij}(t)= 2 d_{ij}(t)$ of the Hermitian decoherence matrix $\mathbf{D}$ are the following:
\begin{eqnarray}  
D_{11}(t) = \frac{1}{2 i \hbar} \left[ \frac{A(t)}{P_e} - \frac{A(t)}{P_g} + \frac{B(t)}{<\psi_{g}|\psi_{e}>}
-\frac{B^*(t)}{<\psi_{e}|\psi_{g}>} \right],\nonumber \\
 \label{D11}
\\
D_{12}(t) = \frac{1}{2 \hbar} \left[ \frac{A(t)}{P_e} + \frac{A(t)}{P_g} - \frac{B(t)}{<\psi_{g}|\psi_{e}>}
-\frac{B^*(t)}{<\psi_{e}|\psi_{g}>} \right],\nonumber \\ 
\label{D12}
\\
D_{21}(t) = D^*_{12}(t),\nonumber \\ 
\label{D1221}
\\
D_{22}(t) = \frac{1}{2 i \hbar} \left[ \frac{A(t)}{P_e} - \frac{A(t)}{P_g} - \frac{B(t)}{<\psi_{g}|\psi_{e}>}
+\frac{B^*(t)}{<\psi_{e}|\psi_{g}>} \right],\nonumber \\ 
\label{D22}
\\
D_{33}(t) = -i \frac{A(t)}{2\hbar} (P_g-P_e) \left[ \frac{1}{|<\psi_{g}|\psi_{e}>|^2} 
- \frac{1}{P_g P_e} \right],\nonumber \\
\label{D33}
\\
D_{13}= D^*_{31} = 0  ~,~ D_{23}= D^*_{32} =0,\nonumber \\
\label{Dij} 
\end{eqnarray}
with $A(t)$ and  $B(t)$ given by Eqs.~(\ref{At}) and (\ref{Bt}). Let us remark that the elements of the
decoherence matrix
are finite as long as $P_g, P_e \ne 0$, and $<\psi_{g}|\psi_{e}> \ne 0$. These conditions
are equally required in order to obtain finite values for the canonical decoherence rates,
 and in the following we will suppose them fulfilled. 

The canonical decoherence rates $\{ \gamma_i(t)\}_{i=1,2,3} $, obtained as eigenvalues 
of the decoherence matrix with elements $D_{ij}(t)$, are
\begin{eqnarray} 
\gamma_{1,2}(t) = \frac{1}{2 P_g P_e} \frac{dP_g}{dt} (P_g-P_e)\nonumber \\ 
\pm
\sqrt{ \left( \frac{1}{2 P_g P_e} \frac{dP_g}{dt} \right)^2  + 
 \frac{1}{|<\psi_{g}|\psi_{e}>|^2} \left| \frac{d <\psi_{g}|\psi_{e}> }{dt} \right|^2 },\nonumber \\
 \label{gama12}
\\ \nonumber \\
\gamma_{3}(t) = \frac{1}{2} \frac{dP_g}{dt} (P_g-P_e) 
 \left[ \frac{1}{|<\psi_{g}|\psi_{e}>|^2} - \frac{1}{P_g P_e} \right]. \nonumber \\
 \label{gama3} 
\end{eqnarray} 

The canonical form of the master equation appears through the diagonalization of the decoherence matrix
(see Eq.~(\ref{diagonalD})).
We deduce the unitary matrix $(U)$ formed by the eigenvectors of the decoherence matrix $(D_{ij})$ as being
\begin{eqnarray}
\label{matrixU}
(U) =
\left(\begin{array}{ccc}
 n_1  &  n_2 & 0 \\ \\
n_1 \frac{\gamma_1 - D_{11}}{D_{12}}  & n_2 \frac{\gamma_2 - D_{11}}{D_{12}} & 0  \\ \\
 0  & 0  &  1
 \end{array} \right),
\end{eqnarray}
with $\gamma_{1,2}$ being the decoherence rates given in Eq.~(\ref{gama12}) and $D_{11}$, $D_{12}$, 
$D_{22}$ being the elements of the decoherence matrix shown in
Eqs.~(\ref{D11} - \ref{D22}). $n_1$ and $n_2$ are real normalization factors (with $n_1^2 + n_2^2 =1$) given by the expressions:
\begin{eqnarray}  
 n_1^2=\frac{\gamma_1 - D_{22}}{\gamma_1 - \gamma_2}  ~;~ 
 n_2^2=\frac{ D_{22} - \gamma_2}{\gamma_1 - \gamma_2}.
 \end{eqnarray}

The time dependent decoherence operators $\{ L_i(t) \}_{i=1,2,3}$, corresponding to orthogonal decoherence channels are obtained using Eq.~(\ref{Lkt}) as
\begin{eqnarray} 
\label{L1t}
L_1(t) = \frac{n_1}{\sqrt{2}} (\sigma_1 + \frac{\gamma_1 - D_{11}}{D_{12}} \sigma_2), \\ \nonumber \\
\label{L2t}
L_2(t) = \frac{n_2}{\sqrt{2}} (\sigma_1 + \frac{\gamma_2 - D_{11}}{D_{12}} \sigma_2), \\ \nonumber \\ 
L_3(t) = \frac{1}{\sqrt{2}} \sigma_3.
\label{L3t}
\end{eqnarray}

Finally, we obtain the canonical form for the master equation of 
the reduced electronic density operator $\hat{\rho}_{el}(t)$ (\ref{matdensel2}):
\begin{eqnarray}
 \frac{d \hat{\rho}_{el}}{dt}= -\frac{i}{\hbar} [ H(t), \hat{\rho}_{el}(t)]\nonumber \\ 
+ \sum_{i=1}^{3}   \gamma_i (t) [ L_i (t) \hat{\rho}_{el} L^+_i (t) - 
\frac{1}{2} \{ L^+_i (t) L_i (t), \hat{\rho}_{el} \} ],
\label{canoneqeldens}
\end{eqnarray}
with the operator $H(t)$ having the matrix (\ref{Hmatrix}), the decoherence rates $\{ \gamma_i(t)\}_{i=1,2,3} $
given in Eqs.~(\ref{gama12}, \ref{gama3}), and the decoherence operators $\{ L_i(t) \}_{i=1,2,3}$ determined by 
Eqs.~(\ref{L1t}-\ref{L3t}).

The sum of the canonical decoherence rates is the trace of the decoherence matrix given by Eqs.~(\ref{D11} - \ref{D33}):
\begin{eqnarray}
\sum_i \gamma_i(t) = \text{Tr}[\mathbf{D}(t)] \nonumber \\
 = \frac{1}{2} \frac{dP_g}{dt} (P_g-P_e) \left[ \frac{1}{P_g P_e} + \frac{1}{|<\psi_{g}|\psi_{e}>|^2} \right],
\label{sumrates}
\end{eqnarray}
becoming zero at instants $t$ for which $dP_g/dt=0$ or $P_g(t)=P_e(t)$.

The Bloch volume of the accessible states, obtained with Eq.~(\ref{Blochvol}), is  
\begin{equation}
\mathcal{V}(t) = \mathcal{V}(t_0)\exp  \left[ - 2 \int_{t_0}^t ds \sum_i \gamma_i(s) \right].
\label{blochvol2}
\end{equation}
As already discussed, if the sum $\sum_i \gamma_i(t)$ of the canonical decoherence rates is negative,
the Bloch volume of the accessible states increases,
witnessing non-Markovianity. Therefore,  a first indication on the non-Markovian behavior 
is given by Eq.~(\ref{sumrates}) which 
shows that a growth of the Bloch volume, $\mathcal{V}(t) > \mathcal{V}(t_0)$, 
appears if $\frac{dP_g}{dt} (P_g-P_e) <0$. 

The normalization condition  $P_g(t) + P_e(t) =1$ implies
\begin{equation}
\frac{dP_g}{dt} (P_g-P_e) = - \frac{d}{dt} ( P_g P_e ).
\label{dtPgPe}
\end{equation}
Therefore, the condition to have $\sum_i \gamma_i(t) < 0$, leading to a growth of the  Bloch volume, can also be expressed as
$\frac{d}{dt}( P_g P_e )>0$.

\subsection{\label{sec:nonmdecohrates} Decoherence rates and canonical measures of non-Markovianity for the electronic system}

Let us analyze the signs of the decoherence rates $\gamma_i(t)$ given by Eqs.~(\ref{gama12},\ref{gama3}). Since 
$P_g P_e$ $\ge$ $|<\psi_{g}|\psi_{e}>|^2$, and with
Eq.~(\ref{dtPgPe}), it appears that the sign of $\gamma_{3}(t)$ depends on the time evolution
of the electronic populations $P_g (t), P_e(t)$ as follows:
\begin{eqnarray}
\text{sgn}[\gamma_{3}(t)] = \text{sgn} \left[ \frac{dP_g}{dt} (P_g-P_e) \right]
= - \text{sgn} \left[ \frac{d}{dt} ( P_g P_e ) \right].\nonumber \\
\label{signgama3}
\end{eqnarray}

On the other hand, Eq.~(\ref{gama12}) can be written as
\begin{eqnarray} 
\gamma_{1,2}(t) = \frac{1}{2P_g P_e } \left| \frac{dP_g}{dt} \right|\nonumber \\
\times
 \left\{ \text{sgn} \left[ \frac{dP_g}{dt} (P_g-P_e) \right] |P_g-P_e|
 \pm 
\sqrt{ 1  + r^2(t)} \right\},\nonumber \\
\label{gama12n}
\end{eqnarray} 
with
\begin{equation}
r^2(t)= \frac {4 P^2_g P^2_e }{( dP_g/dt)^2}
 \frac{| d <\psi_{g}|\psi_{e}> / dt |^2}{|<\psi_{g}|\psi_{e}>|^2}. 
\end{equation}
Taking into account that $0$ $\le$ $|P_g-P_e|$ $\le$  $1$, it becomes obvious that 
$\gamma_1(t)$ is always positive, and $\gamma_2(t)$ is always negative:
\begin{eqnarray}
\gamma_1(t) > 0  ~;  \gamma_2(t) < 0.
\end{eqnarray} 

Consequently, we will distinguish four cases:

(i) If $\frac{dP_g}{dt} (P_g-P_e)$ $>0$, or equivalently, $\frac{d}{dt}( P_g P_e )<0$, there is {\it one negative decoherence rate}, $\gamma_2(t) < 0$, and the non-Markov
 index defined by Eq.~(\ref{nindex}) is $n(t)=1$. Eq.~(\ref{sumrates}) shows that the sum of the decoherence rates is positive,  $\sum_i \gamma_i(t) >0$, leading to a diminution of the Bloch volume. 

The non-Markovianity measure obtained with 
Eqs.~(\ref{fkt},\ref{sumfkt}) is $f(t)=f_2(t)$$=|\gamma_{2}(t)|$.  Using Eq.~(\ref{gama12n}) we find
\begin{equation}
f(t) = \frac{1}{2P_g P_e } \left| \frac{dP_g}{dt} \right| \left[ \sqrt{ 1  + r^2(t)} - |P_g-P_e| \right].
\end{equation}

(ii) If $\frac{dP_g}{dt} (P_g-P_e)$ $<0$, or equivalently, $\frac{d}{dt}( P_g P_e )>0$, there are {\it two negative decoherence rates}, $\gamma_2(t) < 0$ and $\gamma_3(t) < 0$. The dimension of the space of non-Markovian evolution, given by the non-Markov index \cite{hall2014}, becomes $n(t)=2$. The non-Markovianity measure is obtained from the negative decoherence rates using Eqs.~(\ref{fkt}) and (\ref{sumfkt}), as $f(t)=f_2(t)+f_3(t)$$=|\gamma_{2}(t)|+|\gamma_{3}(t)|$. Using Eqs.~(\ref{gama12n},\ref{gama3})
we find
\begin{eqnarray}
f(t) = \frac{1}{2P_g P_e } \left| \frac{dP_g}{dt} \right| \left[ |P_g-P_e| 
+ \sqrt{ 1  + r^2(t)} \right] \nonumber \\
+ \frac{1}{P_g P_e } \frac{d( P_g P_e )}{dt} \frac{ L(t)}{[C_{l_1}( \hat{\rho}_{el})]^2}.
\label{ftenhance}
\end{eqnarray}
In Eq.~(\ref{ftenhance}), $L(t)$ and $C_{l_1}(\hat{\rho}_{el})$ are the linear entropy of 
the electronic-vibrational entanglement and the electronic coherence, respectively, whose 
expressions can be derived from 
Eqs.~(\ref{linentrgen},\ref{l1normgen}) for $N_{el}=2$. 

Moreover, {\it the sum of the decoherence rates is negative}, $\sum_i \gamma_i(t) < 0$, 
which means that
{\it the Bloch volume} of the dynamically accessible states {\it increases} (Eq.~\ref{Blochvol}),
witnessing non-Markovianity. We distinguish this case as indicating {\it enhancement of non-Markovianity}.

(iii) If $P_g(t)=P_e(t)$, the decoherence rates are $\gamma_{3}(t)=0$, and $\gamma_{2}(t)= -\gamma_{1}(t)$.
 The sum of the decoherence rates becomes zero,  $\sum_i \gamma_i(t)=0$.
Using Eq.~(\ref{gama12n}), the non-Markovianity measure $f(t)=|\gamma_{2}(t)|$ becomes
\begin{equation}
f(t) = \frac{1}{2P_g P_e } \left| \frac{dP_g}{dt} \right| \sqrt{ 1  + r^2(t)}. 
\end{equation}

(iv) If $\frac{dP_g}{dt}=0$. This condition corresponds to extrema in the evolution of the 
electronic populations during the pulse, or to constant populations after pulse.
 The decoherence rates become $\gamma_{3}(t)=0$, and $\gamma_{2}(t)= -\gamma_{1}(t)$, with $\sum_i \gamma_i(t)=0$.
Eq.~(\ref{gama12}) gives
\begin{equation} 
\label{gama12cst}
\gamma_{1,2}(t) = \pm \frac{1}{|<\psi_{g}|\psi_{e}>|} \left| \frac{d <\psi_{g}|\psi_{e}> }{dt} \right|,
\end{equation} 
and $f(t)=$$|\gamma_{2}(t)|$.

Let us consider the case of a molecule with constant populations 
in the electronic states $g,e$ (it can be a molecule after the action of a laser pulse): 
$\frac{dP_g}{dt}=0$ for all $t$. Therefore, the Bloch volume of
 the dynamically accessible states remains constant, $\mathcal{V}(t)$$ = \mathcal{V}_0$. 
 For $W(R,t)=0$, Eqs.~(\ref{gama12cst}) and  (\ref{eqS2states}) give an alternative
form of the decoherence  rates as
\begin{eqnarray} 
\gamma_{1,2}(t)=\pm \frac{1}{\hbar} \frac{|<\psi_{g}|V_e(R) - V_g(R) |\psi_{e}>|}{|<\psi_{g}|\psi_{e}>|}. 
 \label{gama12W0}
\end{eqnarray} 
Writing the complex overlap of the vibrational packets as $<\psi_{g}|\psi_{e}>$$=|<\psi_{g}|\psi_{e}>|$
exp($i \alpha(t)$), with $\alpha(t)$ a real function, the non-Markovianity measure
$f(t)=$$|\gamma_{2}(t)|$ obtained using Eq.~(\ref{gama12cst}) becomes
\begin{equation}
f(t) = \sqrt{ \left( \frac{1}{|<\psi_{g}|\psi_{e}>|}\frac{d |<\psi_{g}|\psi_{e}>|}{dt} \right)^2 + 
 \left( \frac{d \alpha}{dt} \right)^2}.
\label{ftisolmol}
\end{equation}
Eq.~(\ref{ftisolmol}) is useful for understanding the relation between $f(t)$ and
 the electronic coherence $|<\psi_{g}|\psi_{e}>|$. It appears that if at an instant $t_m$ one has
 ($\frac{d |<\psi_{g}|\psi_{e}>|}{dt})_{t_m}=0$ (an extremum in the 
time evolution of the coherence), but $|<\psi_{g}|\psi_{e}>|_{t_m} \ne 0$, one obtains a minimum
of the function $f(t)$, which becomes $f(t_m)=| \frac{d \alpha}{dt}|_{t_m}$. On the contrary, 
at an instant $t_M$ for which $|<\psi_{g}|\psi_{e}>|_{t_M} \to 0$ 
(which obviously represents a minimum in the time evolution of the coherence, and therefore
($\frac{d |<\psi_{g}|\psi_{e}>|}{dt})_{t_M}=0$),
the function $f(t)$ has a maximum, becoming 
$f(t_M) $$=\sqrt{ 1 + \left( \frac{d \alpha}{dt} \right)_{t_M}^2}$. 
Eq.~(\ref{ftisolmol}) shows that in a molecule with constant electronic populations, the non-Markovianity measure
$f(t)$ can be seen as a measure of the temporal behavior of the electronic coherence,
having minima when the electronic coherence
has maxima, and attaining maximum values whenever the overlap
of the vibrational packets tends to zero,  $|<\psi_{g}|\psi_{e}>| \to 0$.
At the same time, as we have shown previously \cite{vatasescu2015}, if the electronic populations are constant, the time variations of the coherence $|<\psi_{g}|\psi_{e}>|$ completely determine the temporal evolution of the linear entropy of entanglement $L(t)$ (see Eq.~(\ref{linentr2})), which becomes maximum
when coherence attains a minimum. Therefore, the maxima of the non-Markovianity measure $f(t)$ correspond to maxima of the electronic-vibrational entanglement measured by the linear entropy.

These results make explicit the fundamental non-Markovian character of
 the electronic subsystem evolution.
Indeed, we have shown that one of the decoherence rates is always negative: $\gamma_2(t) < 0$. 
Besides this inherent non-Markovianity,  the character of the electronic evolution becomes strongly
non-Markovian under the condition  $(P_g-P_e)$$dP_g/dt$ $<0$ i.e. $d( P_g P_e )/dt>0$,
which supposes an exchange of population between the electronic channels.
In the following, $d( P_g P_e )/dt$ will be called the non-Markovianity factor.

The condition $(P_g-P_e)$$dP_g/dt$ $<0$ 
implies $\text{sgn}(dP_g/dt)$ $=-\text{sgn}[P_g(t)-P_e(t)]$. Therefore, it appears 
that the non-Markovian character of the dynamics is strengthened when the transfer of population
 between the two electronic channels is such as the larger population decreases
 (i.e., the smaller electronic population increases). This condition, describing an evolution
oriented to
 the equalization of the electronic populations, is in fact a condition indicating the increase of the
electronic-vibrational entanglement, which becomes maximum when the electronic populations are equal \cite{vatasescu2013}. This observation will be developed in the following sections.

\section{\label{sec:nMarkov-entcoh} Connecting non-Markovianity of the electronic evolution with 
electronic-vibrational entanglement and electronic coherence}

We will now analyze enhancement of non-Markovianity, determined by the condition $d( P_g P_e )/dt>0$, in relation to the evolutions of the electronic-vibrational entanglement and the electronic coherence. 
The key observation is that the quantity $P_g(t) P_e(t)$ is connected to measures of 
entanglement and coherence in the molecular system. 

The electronic-vibrational entanglement in the bipartite molecular state $|\Psi_{el,vib}(t) >$ 
given by Eq.~(\ref{2pure-elvib}) can be analyzed
using the von Neumann entropy $S_{vN}(\hat{\rho}_{el}(t))$ or the linear entropy $L(t)$ 
of the reduced density operator $\hat{\rho}_{el}$. In previous works \cite{vatasescu2013,vatasescu2015} we have investigated the results given by these two entanglement measures. Both of them
depend on the temporal behavior of the electronic populations, but only $L(t)$ depends on the electronic coherence. The von Neumann entropy of the
electronic-vibrational entanglement has the following expression \cite{vatasescu2013}:
\begin{equation}
S_{vN}(\hat{\rho}_{el}(t)) = - P_g(t) \log_2 P_g(t)  -  P_e(t)\log_2 P_e(t). 
\label{vonNel}
\end{equation}
For $N_{el}=2$, the linear entropy $L(t)=1- \text{Tr} (\hat{\rho}^2_{el}(t))$ obtained with Eq.~(\ref{linentrgen}) becomes
\begin{equation}
L(t) = 2P_g(t)P_e(t) - 2 |<\psi_{g}(R,t)|\psi_{e}(R,t)>|^2,
\label{linentr2}
\end{equation}
and, with Eq.~(\ref{l1normgen}), the $l_1$ norm measure of the electronic coherence is
\begin{equation}
C_{l_1}( \hat{\rho}_{el})= 2 |<\psi_{g}(R,t)|\psi_{e}(R,t)>|.
\label{cl1norm}
\end{equation}
Therefore, Eq.~(\ref{linentr2}) can be read as a relation between the phenomena of 
electronic-vibrational entanglement, non-Markovianity of the electronic evolution, and electronic coherence. 
Indeed, Eqs.~(\ref{linentr2}) and ~(\ref{cl1norm}) lead to
\begin{equation}
  \frac{d}{dt} [ P_g (t) P_e (t)] = \frac{1}{2} \frac{dL}{dt} +
\frac{1}{2} C_{l_1}( \hat{\rho}_{el}) \frac{d C_{l_1}( \hat{\rho}_{el})}{dt}.
\label{derivLt}
\end{equation}
In the following, Eq.~(\ref{derivLt})  will be used to explore the relations between 
enhancement of non-Markovianity  ($d( P_g P_e )/dt>0$),
increase of entanglement ($dL/dt>0$), and increase of the electronic coherence
 ($dC_{l_1}( \hat{\rho}_{el})/dt>0$).

Expressions of the decoherence rates as functions of $L(t)$ and $C_{l_1}(t)$ can be given. Using Eq.~(\ref{sumrates}), the sum of the decoherence rates becomes
\begin{equation}
\sum_i \gamma_i(t) = - \frac{ d [\ln (P_g P_e)]}{dt}
\frac{ L(t) + [C_{l_1}( \hat{\rho}_{el})]^2} {[C_{l_1}( \hat{\rho}_{el})]^2},
\end{equation}
and, with Eq.~(\ref{gama3}), $\gamma_{3}(t)$ can be written
\begin{equation}
\gamma_{3}(t) = - \frac{ d [\ln (P_g P_e)]}{dt} \frac{ L(t)}{[C_{l_1}( \hat{\rho}_{el})]^2}.
\end{equation}

\begin{table*}
\caption{\label{tab:relations} Connections between the time behavior of the electronic-vibrational entanglement ($dL/dt$), the enhancement of non-Markovianity in the evolution of the electronic subsystem ($d(P_g P_e)/dt>0$), and behaviors of speakable and unspeakable \cite{spekkens16} electronic coherences, measured by $l_1$ norm
 $C_{l_1}(t)$ and skew information ${\cal I_S}(t)$, respectively.}
\begin{ruledtabular}
\begin{tabular}{cccccccc}
&$\frac{dL}{dt}$&$\frac{d (P_g P_e)}{dt}$&&$\frac{d C_{l_1}}{dt}$&&&$\frac {\partial {\cal I_S}  }{ \partial t}$\\  \hline   
&&&&&&& \\   
(1)&$>0$&$<0$&$\Longrightarrow$&$<0$&&&$<0$ \\
 \hline \\
(2)&$>0$&$>0$&$\Longrightarrow$&$>0$, if&&&$>0$, if \\
&&&&$\frac{d( P_g P_e )}{dt}>\frac{1}{2}\frac{dL}{dt}$&&&$\frac{1}{ 2\sqrt{2L}(1+ \sqrt{2L})}\frac{dL}{dt} <\frac{1}{ \sqrt{2L} + L + 2 P_g P_e}\frac{d(P_g P_e)}{dt}
<\frac{1}{C_{l_1}}\frac{d C_{l_1}}{dt} $ \\
&&&&&&& \\
&&&&&&&$<0$, if  \\
&&&&&&&$\frac{1}{ 2\sqrt{2L}(1+ \sqrt{2L})}\frac{dL}{dt} >\frac{1}{ \sqrt{2L} + L + 2 P_g P_e}\frac{d(P_g P_e)}{dt}
>\frac{1}{C_{l_1}}\frac{d C_{l_1}}{dt} $  \\
&&&&&&&\\
&&&&&&&\\
&&&&$<0$, if&&&$<0$\\
&&&&$\frac{d( P_g P_e )}{dt}<\frac{1}{2}\frac{dL}{dt}$&&& \\
&&&&&&& \\ \hline 
&&&&&&&\\
(3)&$<0$&$>0$&$\Longrightarrow$&$>0$&&&$>0$ \\
 \hline 
&&&&&&& \\
(4)&$<0$&$<0$&$\Longrightarrow$&$<0$, if&&&$>0$, if \\
&&&&$-\frac{1}{2}\frac{dL}{dt} < -\frac{d( P_g P_e )}{dt}$&&&$-\frac{1}{ 2\sqrt{2L}(1+ \sqrt{2L})}\frac{dL}{dt} >-\frac{1}{ \sqrt{2L} + L + 2 P_g P_e}\frac{d(P_g P_e)}{dt}
>-\frac{1}{C_{l_1}}\frac{d C_{l_1}}{dt} $ \\
&&&&&&& \\
&&&&&&&$<0$, if \\
&&&&&&&$-\frac{1}{ 2\sqrt{2L}(1+ \sqrt{2L})}\frac{dL}{dt} <-\frac{1}{ \sqrt{2L} + L + 2 P_g P_e}\frac{d(P_g P_e)}{dt}
<-\frac{1}{C_{l_1}}\frac{d C_{l_1}}{dt} $ \\
&&&&&&&\\
&&&&&&&\\
&&&&$>0$, if&&&$>0$\\
&&&&$-\frac{1}{2}\frac{dL}{dt} > -\frac{d( P_g P_e )}{dt}$&&& \\
\end{tabular}
\end{ruledtabular}
\end{table*}

Besides the $l_1$ norm measure of the electronic coherence, $C_{l_1}( \hat{\rho}_{el})$,
we shall use the Wigner-Yanase skew information 
${\cal I_S} (\hat{\rho}_{el}, \hat{H}_{el})$$= -\frac{1}{2}
\text{Tr}_{el}[\sqrt{\hat{\rho}_{el}},\hat{H}_{el}]^2 $ for the electronic state $\hat{\rho}_{el}$, 
with respect to the electronic Hamiltonian $\hat{H}_{el}$, to additionally characterize electronic subsystem coherence \cite{vatasescu2015,*vatasescu2016}. The skew information
${\cal I_S}$ is a measure of coherence as asymmetry relative to a group of translations
 \cite{spekkens16,marvian16,qcohRMP17},
quantifying the coherence of a state with respect to a certain Hamiltonian eigenbasis. This
notion of coherence was termed unspeakable \cite{spekkens16}, to show 
its structural relation to the eigenvalues of the observable which defines the basis relative to which coherence is defined \footnote{The term ``unspeakable coherence`` derives from the syntagm ``unspeakable information``, which designates an information which can only be encoded in certain
degrees of freedom \cite{spekkens16}. Also following Ref.~\cite{spekkens16}, ``speakable information is information for which the means of encoding is irrelevant`` (p. 2). The term ``unspeakable coherence`` refers to the notion of ``coherence as asymmetry``, and the term ``speakable coherence`` is applied
to the concept of coherence defined in the recently developed framework of resource theories of quantum coherence \cite{spekkens16,marvian16,qcohRMP17}. It was shown that measures of coherence are a subset of measures of asymmetry \cite{marvian16}.}. It is a notion of coherence closely
related to the context of quantum speed limits \cite{marvian16,qcohRMP17}. 
In particular, ${\cal I_S} (\hat{\rho}_{el}, \hat{H}_{el})$
characterizes the coherence of the reduced electronic state $\hat{\rho}_{el}$ relative to the eigenbasis 
$\{ |g>,|e> \}$ of the
electronic Hamiltonian $\hat{H}_{el}$, whose eigenvalues are the electronic potentials $V_{g}(R),V_{e}(R)$.
The skew information ${\cal I_S} (\hat{\rho}_{el}, \hat{H}_{el})$ has the following expression \cite{vatasescu2015,*vatasescu2016}:
\begin{eqnarray}
{\cal I_S} (\hat{\rho}_{el}, \hat{H}_{el}) = [V_g(R) - V_e(R)]^2 
\frac{|<\psi_{g}(R,t)|\psi_{e}(R,t)>|^2}{ 1+ \sqrt{2L(t)} }. \nonumber \\ 
\label{skewinfel2}
\end{eqnarray}
${\cal I_S} (\hat{\rho}_{el}, \hat{H}_{el})$$={\cal I_S}(R,t)$ appears as a product between a  function of the internuclear distance $R$
(depending on the electronic potentials difference at given $R$) and a function of time $t$, 
a factorization which reflects the BO approximation. It can be said that ${\cal I_S}(R,t)$ is a measure of the unspeakable electronic coherence which characterizes the reduced electronic state $\hat{\rho}_{el}$ at a given internuclear distance $R$. Let us observe that the time behavior of ${\cal I_S}$ 
is determined by the time evolutions of the  electronic coherence $C_{l_1}(t)$ and the linear entropy of entanglement $L(t)$. Our aim is to investigate non-Markovian behavior in relation to various quantum correlations in the molecular system, and we find it useful to also examine this  measure of correlations,
which combines coherence and entanglement.

Eqs.~(\ref{skewinfel2}) and ~(\ref{cl1norm}) determine the relation between 
the time variations of the electronic coherences
${\cal I_S}(R,t)$, $C_{l_1}(t)$, and of the linear entropy of entanglement $L(t)$:
\begin{eqnarray}
\frac{1}{{\cal I_S}} \frac {\partial {\cal I_S}}{\partial t} = \frac{2}{C_{l_1}} \frac{d C_{l_1}}{dt}
 - 
\frac{ 1 }{ \sqrt{2L}(1+ \sqrt{2L})}  \frac{dL}{dt}. 
\label{dIsdt}
\end{eqnarray}

We shall analyze the condition $d( P_g P_e )/dt >0$ of enhanced non-Markovian behavior in the electronic evolution 
in connection to the time behaviors of entanglement and the 
two kinds of electronic coherence ("speakable", quantified by the $l_1$ norm $C_{l_1}$, 
 and "unspeakable" \cite{spekkens16}, quantified by the skew information ${\cal I_S}$).  
Eqs.~(\ref{dIsdt}) and (\ref{linentr2}) give:
\begin{eqnarray}
\frac{d(P_g P_e)}{dt} = \frac{ \sqrt{2L} + L + 2 P_g P_e }{2 \sqrt{2L} (1+ \sqrt{2L})} 
 \left( \frac{dL}{dt} \right)
+\frac{C^2_{l_1}}{ 4 {\cal I_S}} \left( \frac { \partial {\cal I_S}  }{ \partial t} \right),
  \nonumber \\
\label{nMskewent}
\end{eqnarray}
\begin{eqnarray}
\frac{d(P_g P_e)}{dt} = \frac{ \sqrt{2L} + L + 2 P_g P_e  }{C_{l_1}}
  \left( \frac{d C_{l_1}}{dt} \right) \nonumber \\
-
\frac{ \sqrt{2L} (1+ \sqrt{2L})^2 }{2 {\cal I_S}}
\left( \frac {\partial {\cal I_S}  }{ \partial t} \right). 
\label{nMskewcoh}
\end{eqnarray}

Table~\ref{tab:relations} systematizes the relations between enhancement of non-Markovianity
($d(P_g P_e )/dt>0$) and the dynamics of the quantum correlations measured using
$L(t)$, $C_{l_1}(t)$, and skew information ${\cal I_S} (\hat{\rho}_{el}, \hat{H}_{el})$. 
This analysis is performed using
Eqs.~(\ref{derivLt},\ref{dIsdt},{\ref{nMskewent},\ref{nMskewcoh}). Observing that
 non-Markovian behavior accompanies
 the phenomenon of electronic-vibrational entanglement, we have considered definite signs
for $dL/dt$ and  $d(P_g P_e )/dt$, in order to deduce the compatible behaviors of electronic coherences.
Table~\ref{tab:relations} shows the following relations among phenomena:

 (1) Entanglement growth ($dL/dt>0$) accompanied by diminution of non-Markovianity  
($d (P_g P_e)/dt<0$) has to be associated with a decrease of both electronic coherences 
($C_{l_1}$ and ${\cal I_S}$). 

 (2) When both entanglement and non-Markovianity increase ($dL/dt>0$, $d (P_g P_e)/dt>0$),
the electronic coherence $C_{l_1}$ may either increase
 (if $\frac{d( P_g P_e )}{dt}>\frac{1}{2}\frac{dL}{dt}$), 
or decrease (if the opposite relation is true). 
If $d C_{l_1} / dt >0$, the skew information can increase or decrease, 
depending on the hierarchy among
the time behaviors of $L(t)$, $P_g (t)P_e(t)$, and $C_{l_1}(t)$, as it is shown in the fourth column
of the Table~\ref{tab:relations}.  On the contrary, if $d C_{l_1} / dt <0$, 
the skew information can only decrease, $ \partial {\cal I_S} / \partial t <0$.

(3) Decrease of entanglement ($dL/dt<0$) is accompanied by enhanced non-Markovian behavior
($d(P_g P_e)/dt>0$) only if the electronic coherences ($C_{l_1}$ and ${\cal I_S}$) increase.

(4) When both entanglement and non-Markovianity decrease ($dL/dt<0$, $d (P_g P_e)/dt<0$),
the electronic coherence $C_{l_1}$ may either increase or decrease. As in the case (2), we will
have several possibilities, shown in the Table.
 
We observe a notable difference between the cases (2) and (4), with
$dL/dt$, $d(P_g P_e)/dt$ having the same sign, and cases (1) and (3), with them having opposite signs. 
The numerical results presented in Sec.~\ref{sec:molecexamples} will show that cases (2) and (4) represent the rule, and cases (1) and (3) are the exception, because enhanced non-Markovian behavior is deeply connected with increase of entanglement, as already explained in Sec. \ref{sec:nonmdecohrates}.

It is interesting to compare the time behaviors 
of the two electronic coherences: Even if the skew information
has the tendency to follow the $C_{l_1}$ time behavior, its sensitivity to entanglement 
brings cases in which the increase of the electronic coherence $C_{l_1}$ 
is accompanied by the decrease of ${\cal I_S}$, 
or the opposite. The conditions of possibility leading to these situations appear 
in the cases (2) and (4),
specified in Table ~\ref{tab:relations}. 

The aim of this analysis is to gain insight into the meaning of non-Markovianity in relation to
entanglement and coherence. An interesting question would be if the model used here to characterize
non-Markovianity allows us to relate non-Markovian behavior to a backflow of information
 from environment to the system. More specifically, the question is if any of the conditions 
$d(P_g P_e)/dt>0$,  $dL/dt>0$, or $d C_{l_1} / dt >0$ could be related to a flow of information from the 
vibrational environment to the electronic open subsystem. As is well known, Breuer {\it et al.} \cite{breuerlp09}
 identify as an essential feature of non-Markovian behavior the existence
of a reversed flow of information from the environment to the open system, a
"backflow" which is manifested in the growth of distinguishability between quantum states of the open system. 
In the Appendix we show that the trace distance between $\hat{\rho}_{el}(t)$ and a state 
$\hat{\rho}_{el}(t_0)$ with coherence $C_{l_1}(t_0)=0$ is increased when $d( P_g P_e )/dt>0$ and $d C_{l_1} / dt >0$.
In general (see the appendix), the condition $d( P_g P_e )/dt>0$ 
for enhanced non-Markovian behavior participates in the increase of the trace distance 
$D(\hat{\rho}_{el}(t_0),\hat{\rho}_{el}(t))$, contributing with a positive term 
at the rate of change  $dD(\hat{\rho}_{el}(t_0),\hat{\rho}_{el}(t))/ dt$  given by 
Eq.~(\ref{dtdistel}).
Regarding the condition $dL/dt>0$,  Sec. \ref{sec:nonmdecohrates} explained 
that the condition $(P_g-P_e)$$dP_g/dt$ $<0$ indicating
enhanced non-Markovian behavior describes an evolution of the electronic populations 
which increases entanglement. The close bond between the condition  $d( P_g P_e )/dt>0$ and the increase of entanglement ($dL/dt>0$, $dS_{vN}/dt>0$) will appear clearly in the numerical results presented
in Sec.~\ref{sec:molecexamples}. 

This theoretical analysis, grounded on the analytic formulas relating
the non-Markovianity factor $d( P_g P_e )/dt$ with the time behaviors of entanglement and coherence, 
 will be completed 
in Sec.~\ref{sec:molecexamples} with an examination of numerical results for the canonical measures of non-Markovianity
obtained from simulations of the molecular dynamics in a laser-driven molecule.

\section{\label{sec:nMarkov-uncert} Non-Markovianity and quantum uncertainty on the electronic energy}

If $\hat{\rho}_{el,vib}(t)$ is a pure state, the uncertainty on the electronic energy (i.e. the mean square deviation from the average value) is given by \cite{vatasescu2015,*vatasescu2016}
\begin{eqnarray}
(\Delta \hat{H}_{el})^2 = {\cal I_S} (\hat{\rho}_{el,vib}, \hat{H}_{el}\bigotimes \hat{I}_v) \nonumber \\
= [V_g(R) - V_e(R)]^2 P_g (t) P_e (t),
\label{skewelvib2} 
\end{eqnarray}
where ${\cal I_S} (\hat{\rho}_{el,vib}, \hat{H}_{el}\bigotimes \hat{I}_v)$ is the Wigner-Yanase skew information for the molecular state $\hat{\rho}_{el,vib}$ with respect to  the electronic Hamiltonian $\hat{H}_{el}$.  Consequently, enhancement of non-Markovianity in the electronic evolution increases uncertainty on the electronic energy 
(and inversely, growing uncertainty on the electronic energy reflects a non-Markovian behavior in the electronic evolution):
\begin{eqnarray}
\frac{d( P_g P_e )}{dt} >0 \Longleftrightarrow 
\frac{\partial (\Delta \hat{H}_{el})^2  }{\partial t} > 0.
\end{eqnarray}
The Wigner-Yanase skew information ${\cal I_S} (\hat{\rho}_{el}, \hat{H}_{el})$ is also recognized as a measure of 
the quantum uncertainty of $\hat{H}_{el}$ in the state $\hat{\rho}_{el}$ 
\cite{luomathphys05,*luo05,*luo06}. Let us observe that Eq.~(\ref{nMskewcoh}) connects the time behavior of the uncertainty on the electronic energy
in the pure molecular state $\hat{\rho}_{el,vib}(t)$ with behavior of the quantum uncertainty ${\cal I_S} (\hat{\rho}_{el}, \hat{H}_{el})$ in the reduced state $\hat{\rho}_{el}$.

\section{\label{sec:molecexamples} Non-Markovian dynamics of the electronic subsystem in a laser-driven molecule: analysis from simulations of molecular dynamics}

\begin{figure} 
\includegraphics[width=0.8\columnwidth]{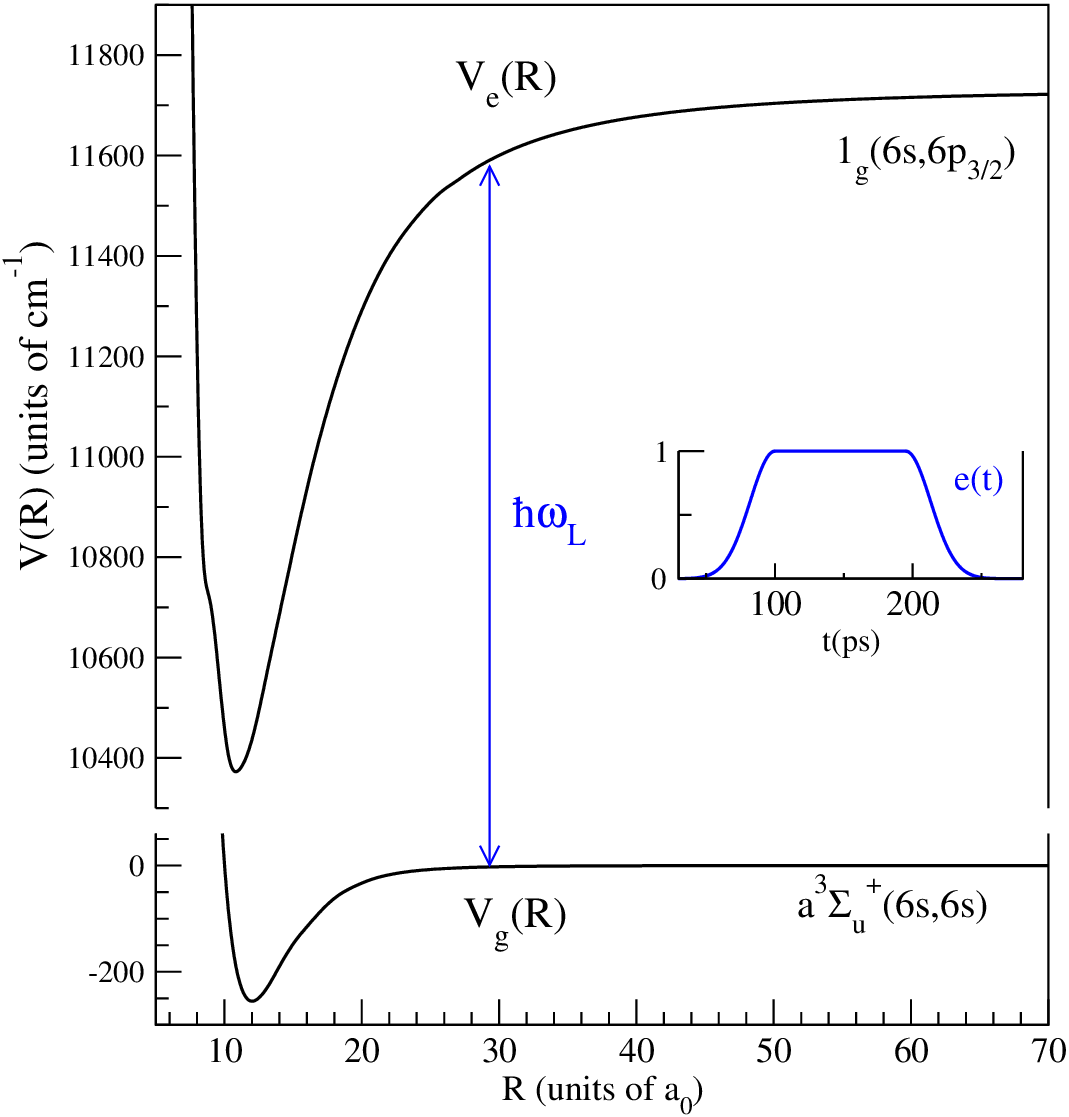}%
 \caption{\label{fig1_pot} (Color online) $a^3\Sigma_{u}^{+}(6s,6s)$ and
  $1_g(6s,6p_{3/2})$ electronic
potentials of Cs$_2$, coupled at a internuclear distance of about  $R_c \approx 29\ a_0$ by a pulse with frequency $\omega_L/2\pi$ and envelope $e(t)$ shown
in the inset. The energy origin is taken to be the
dissociation limit $E_{6s+6s}=0$ of the $a^3\Sigma_{u}^{+}(6s,6s)$
potential.}
\end{figure}

This section will present results obtained from the simulation of the intramolecular dynamics
for a diatomic molecule which is under the action of a laser pulse coupling two electronic states.
Non-Markovian behavior of the electronic subsystem is characterized using the 
canonical measures of non-Markovianity $f(t)$ and $F(t_1,t_2)$$=\int_{t_1}^{t_2} f(t) dt$, calculated
using the equations established in Sec. \ref{sec:nonmdecohrates}. We will also examine the time behavior of the Bloch volume $\mathcal{V}(t)$ of the accessible states, obtained  using Eqs.~(\ref{sumrates},\ref{blochvol2}), as well as the dynamics of the electronic-vibrational entanglement and the electronic coherence in the molecule.
Non-Markovian behavior during time evolution will be connected with the
dynamics of quantum correlations.

As a model system, we consider the Cs$_2$ molecule in which 
the electronic states $g=a^3\Sigma_{u}^{+}(6s,6s)$ and $e=1_g(6s,6p_{3/2})$ 
are coupled by a laser pulse. In previous works \cite{vatasescu01,vatasescu09,vatasescu2013} ,
we have analyzed the vibrational dynamics in these electronic potentials for various conditions
of coupling, and we shall refer to these works for details of the molecular model, 
 including definitions of the characteristic times of dynamics, such as vibrational and Rabi periods. 

\begin{figure}
\includegraphics[width=0.95\columnwidth]{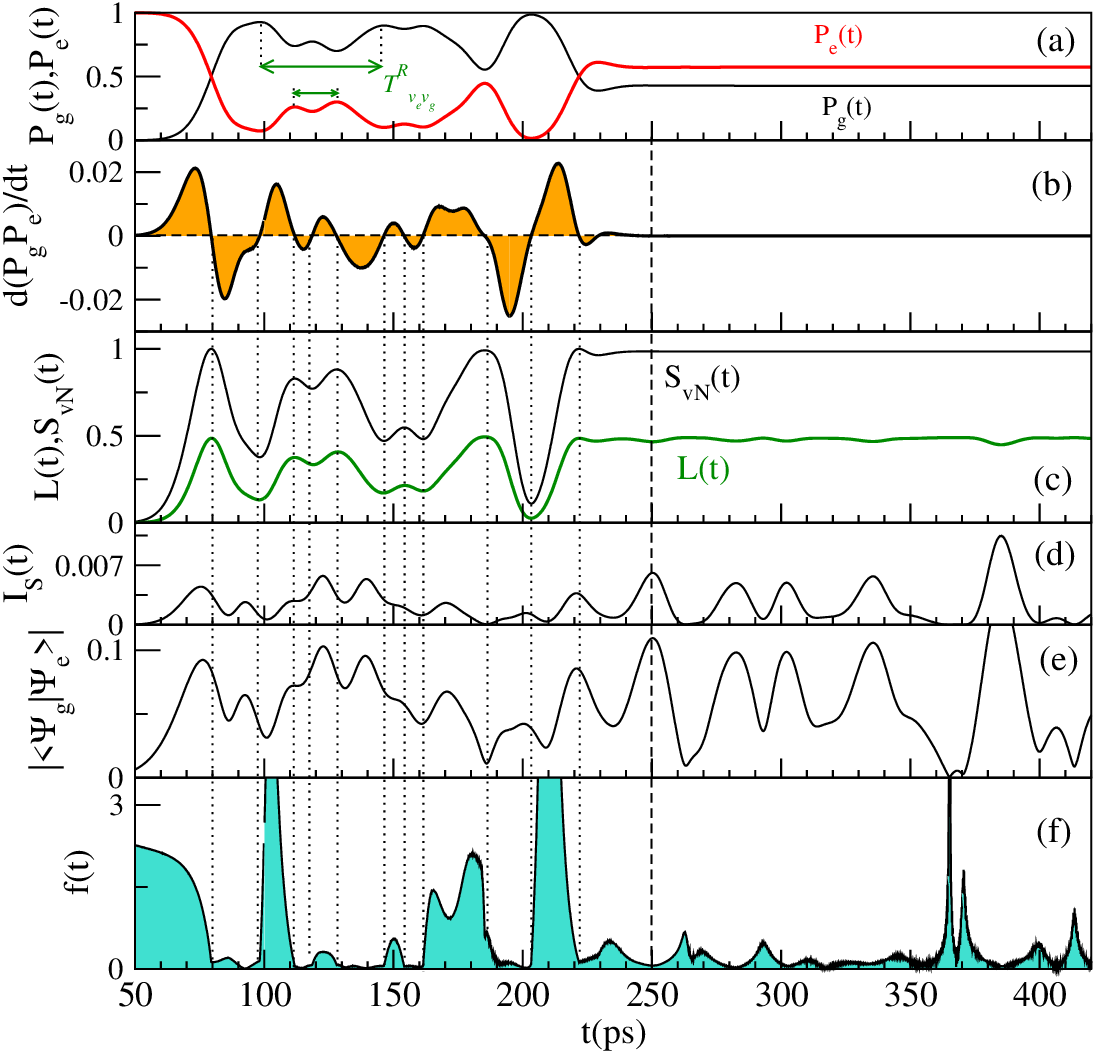}%
 \caption{\label{fig2_WL} (Color online) Results characterizing the vibrational dynamics in the
 electronic potentials $g=a^3\Sigma_{u}^{+}$ and $e=1_g$ of Cs$_2$ coupled by a pulse with envelope
$e(t)$ (Fig.~\ref{fig1_pot}), for a coupling strength $W_L=3.29$ cm$^{-1}$. Time evolutions during
the pulse ($t<250$ ps) and after pulse ($t>250$ ps) are both shown.
(a) Time evolution of the populations $P_g(t)$ and $P_e(t)$
(two specific Rabi periods $T^R_{v_e,v_g}$, of $47.4$ ps and $16.5$ ps, are marked). 
(b) Time evolution of the
 "non-Markovianity factor"  $d( P_g P_e )/dt$ (non-Markovianity is enhanced if $d( P_g P_e )/dt>0$).
(c) Time evolutions of the linear entropy $L(t)$ and von Neumann entropy
$S_{vN}(t)$ of the electronic-vibrational entanglement.
(d) Time evolution of the skew information $I_S(t)={\cal I_S}(R,t)/[\Delta V(R)]^2$.
(e) Time evolution of the electronic coherence $C_{l_1}(t)/2=|<\psi_{g}(t)|\psi_{e}(t)>|$.
(f) Non-Markovianity measure $f(t)$. The filled surface shows the integral $\int f(t) dt$.}
\end{figure}

Let us suppose the electronic states $g=a^3\Sigma_{u}^{+}(6s,6s)$ and $e=1_g(6s,6p_{3/2})$ coupled by 
an electric field with temporal amplitude 
${\cal{E}}(t)={\cal {E}}_0 e(t) \cos \omega_Lt$.
The field amplitude ${\cal {E}}_0=\sqrt{2I/c\epsilon_0}$ depends on the laser intensity $I$, 
$e(t)$ is the temporal envelope of the pulse, and  $\omega_L/2\pi$
is the frequency of the field, such as the photon energy $\hbar \omega_L$ couples the electronic potentials
$V_g(R)$ and $V_e(R)$ at a internuclear distance of about  $R_c \approx 29\ a_0$, as it is shown in 
 Fig.~\ref{fig1_pot}. 
Using the rotating wave approximation with the frequency $\omega_L/2\pi$, and a transformation of the radial
wave functions with appropriate phase factors, one obtains the typical
Eq.~(\ref{eqS2states}) for the vibrational wave packets $\psi_{g}(R,t)$ and $\psi_{e}(R,t)$ whose dynamics takes place 
in the diabatic electronic potentials crossing in $R_c$ \cite{vatasescu01}. The coupling between
 the electronic channels is $W(t)=W_L e(t)$, with the strength
$W_L= - \frac {1}{2}{\cal {E}}_0 D_{ge}^{\vec{e_L}}$, where 
 $D_{ge}^{\vec{e_L}}$  is the transition dipole moment between the ground $g$ and the excited $e$ 
electronic states, for a polarization ${\vec{e_L}}$ of the electric field \cite{vatasescu01}. Here 
the $R$-dependence of the transition dipole moment is neglected, and
 several coupling strengths $W_L$ are considered, for the same pulse envelope $e(t)$ (represented in Fig.~\ref{fig1_pot}).

\begin{figure}
\includegraphics[width=0.95\columnwidth]{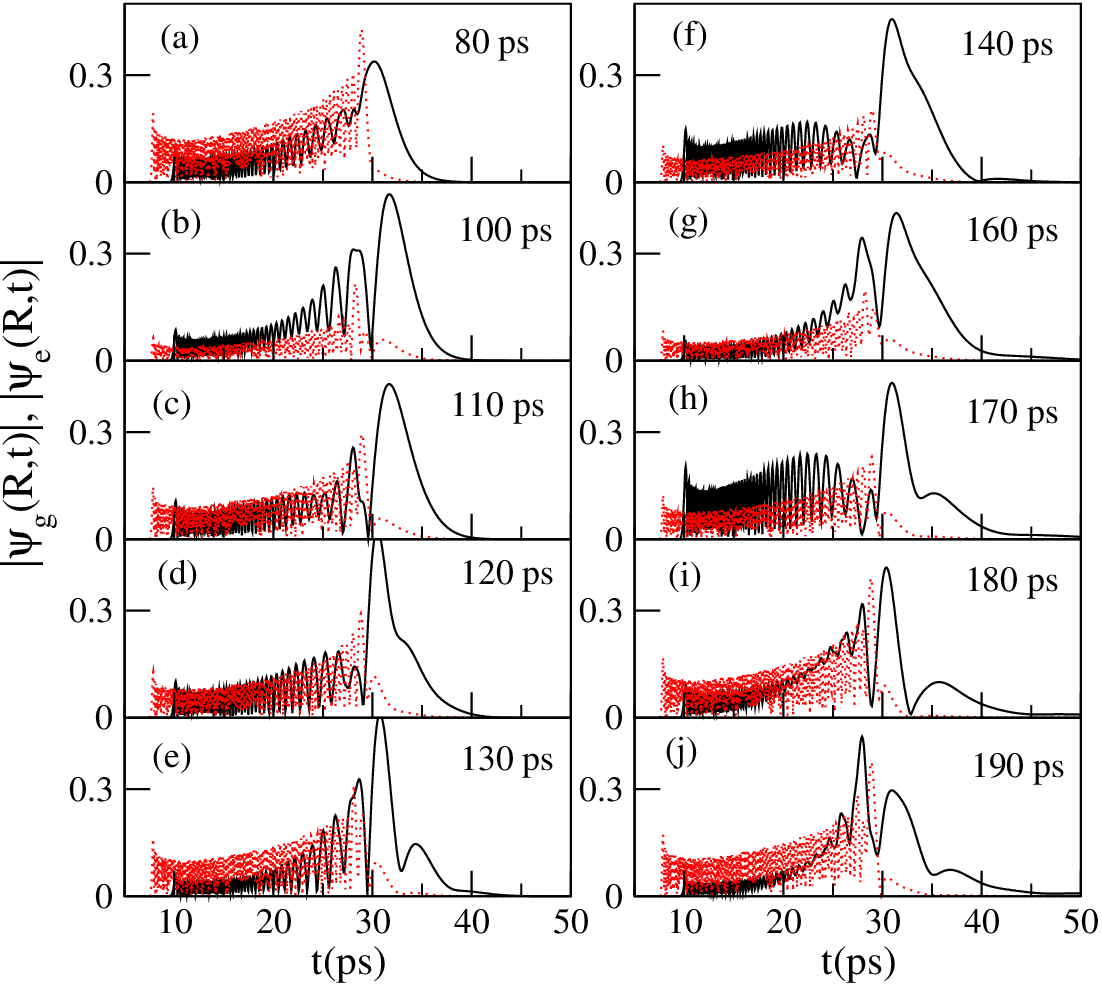}%
 \caption{\label{fig3_wf1WL} (Color online) Time evolution (80 - 190 ps) of the vibrational wave packets 
$|\Psi_g(R,t)|$ (full line) and $|\Psi_e(R,t)|$ (dotted line) in $g=a^3\Sigma_{u}^{+}(6s,6s)$ and $e=1_g(6s,6p_{3/2})$ electronic potentials coupled by a pulse with envelope
$e(t)$ (Fig.~\ref{fig1_pot}), for a coupling strength $W_L=3.29$ cm$^{-1}$.}
\end{figure}

The intramolecular dynamics is obtained using
Eq.~(\ref{eqS2states}), which is solved numerically by propagating in time
 an initial wave function (here the initial state is the vibrational eigenstate with $v_e=142$
of the $1_g(6s,6p_{3/2})$ potential) on a spatial grid with length  $L_R$. 
The Mapped Sine Grid (MSG) method \cite{elianeepjd04, willner04} is used
 to represent the radial dependence of the
 wave packets, and the time
propagation uses the Chebychev expansion of the evolution operator \cite{kosloff94,kosloff96}.
The electronic populations $P_{g}(t)$, $P_{e}(t)$  are calculated from the vibrational wave packets
 as $P_{g,e}(t) = \int^{L_R} | \Psi_{g,e}(R',t) |^{2} dR'$, and the electronic coherence
(\ref{cl1norm}) is obtained from the overlap of the vibrational wave packets calculated on the spatial grid:
 $<\psi_{g}(t)|\psi_{e}(t)>$$=\int^{L_R} \Psi^*_{g}(R',t)\Psi_{e}(R',t)dR'$.
These results are used to calculate the canonical decoherence rates and measures of non-Markovianity, as well as the entropies of the electronic-vibrational entanglement and the skew information.

We begin by analyzing dynamics for a coupling strength $W_L=3.29$ cm$^{-1}$
(corresponding to a pulse intensity $I \approx 2.7$ MW/cm$^2$ for a linear polarization vector
  ${\vec{e_L}}$  \cite{vatasescu99}), for which the results are given in Figs.~\ref{fig2_WL},\ref{fig3_wf1WL} and \ref{fig4_wf2WL}. Fig.~\ref{fig2_WL} shows the time evolutions of several significant quantities:
electronic populations $P_{g}(t)$, $P_{e}(t)$, "non-Markovianity factor"  $d( P_g P_e )/dt$, 
entropies $L(t)$ and $S_{vN}(t)$ of the electronic-vibrational entanglement, 
 electronic coherence $C_{l_1}(t)$ and skew information
$I_S(t)$, as well as the non-Markovianity measure $f(t)$. The vertical dotted lines in the figure 
help us to observe the correlations between the temporal variations of all these properties. 
Figs.~\ref{fig3_wf1WL} and \ref{fig4_wf2WL}
show  the time evolution of the vibrational wave packets $|\Psi_g(R,t)|$ 
and $|\Psi_e(R,t)|$, during the pulse and after pulse. 

The pulse, which operates from 50 to 250 ps (see the envelope
$e(t)$ in Fig.~\ref{fig1_pot}), couples the two electronic states activating a vibrational dynamics which involves several vibrational levels
of each surface, with vibrational periods of about 11 ps
in the $1_g$ electronic potential (the vibrational levels $v_e=140$ up to $143$ are implied), 
and between 33 and more than  100 ps in the $a^3\Sigma_{u}^{+}$ potential (corresponding mainly to the vibrational levels from $v_g=43$ up to $49$). 
The pulse produces a rich vibrational dynamics, 
implying  transfer of population between the electronic states, inversion of population, and beats with various Rabi periods $T^R_{v_e,v_g}$ \cite{vatasescu09} between the populated vibrational levels of the excited and ground states. These phenomena are visible in Fig.~\ref{fig2_WL}(a), where typical Rabi periods can be identified, such as $T^R_{v_e,v_g}=47.4$ ps (between  $v_e=142$ of $1_g$  and $v_g=47$ of $a^3\Sigma_{u}^{+}$) and $T^R_{v_e,v_g}=16.5$ ps (between  $v_e=142$, $v_g=45$).
The time evolution of the wave packets in Figs.~\ref{fig3_wf1WL} and \ref{fig4_wf2WL} allows us to 
observe the relation between the population transfer between electronic channels and the vibrational motion in the potential wells. Let us briefly decipher the dynamics from these results.
The pulse begins by transferring  electronic population from $e=1_g$ state ($P_e(0)=1$) to
$g=a^3\Sigma_{u}^{+}$ state, the populations becoming equals at about 80 ps. This process, taking place
from 50 to 80 ps, increases entanglement (Fig.~\ref{fig2_WL}(c)), and is associated with a strong 
non-Markovian behavior (Fig.~\ref{fig2_WL}(f)). After 80 ps, $P_g(t) > P_e(t)$, and the population transfer
from $e$ to $g$ continues with the diminution of the entanglement and the non-Markovianity measure $f(t)$. 
The inversion of population is almost completed at 100 ps, and the transfer is inverted, producing a non-Markovianity maximum between 100 and 110 ps (Fig.~\ref{fig2_WL}(f)), followed by stabilization of populations with small Rabi beatings between 110 and 130 ps. The vibrational motion inside the $a^3\Sigma_{u}^{+}$ potential empties
the transfer zone located around the crossing point $R_c \approx 29\ a_0$ 
(see Fig.~\ref{fig3_wf1WL}(f), t=140 ps), therefore between 130 and 140 ps 
the population is transferred from 
$1_g$ to $a^3\Sigma_{u}^{+}$, diminishing the entanglement and the function $f(t)$.
Between 160 and 190 ps, the pulse again transfers again population  from the $g=a^3\Sigma_{u}^{+}$ state
to the $e=1_g$ state, increasing the entanglement and the non-Markovianity function $f(t)$
(this process is temporarily stopped around 170 ps by the vibration of the $g=a^3\Sigma_{u}^{+}$ packet, as
 shown in Fig.~\ref{fig3_wf1WL}(h)). Finally, before the end of the pulse, the massive transfer
of population from the $g=a^3\Sigma_{u}^{+}$ state
to the $e=1_g$ state, between 200 and 220 ps, increases the entanglement and has
a notable non-Markovian character (see Figs.~\ref{fig2_WL}(a,c,f) and  \ref{fig4_wf2WL}(a-c)).
\begin{figure}
\includegraphics[width=0.95\columnwidth]{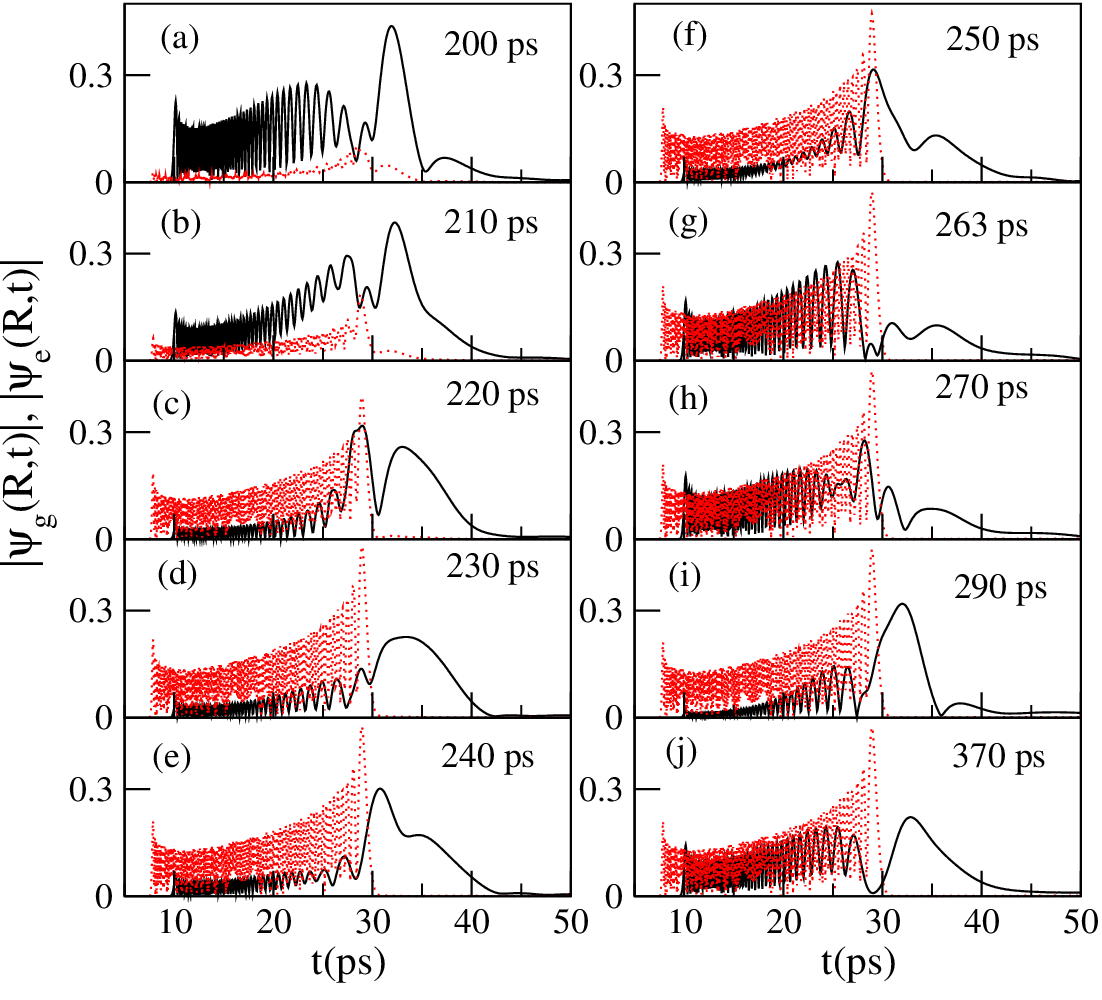}%
 \caption{\label{fig4_wf2WL} (Color online) Continuation of Fig.~\ref{fig3_wf1WL}: time evolution (200 - 370 ps) of the vibrational wave packets 
$|\Psi_g(R,t)|$ (full line) and $|\Psi_e(R,t)|$ (dotted line) for a coupling $W_L=3.29$ cm$^{-1}$. (a-e) Time evolution during the pulse.
(f-j) Time evolution after pulse.}
\end{figure}

\begin{figure}
\includegraphics[width=0.95\columnwidth]{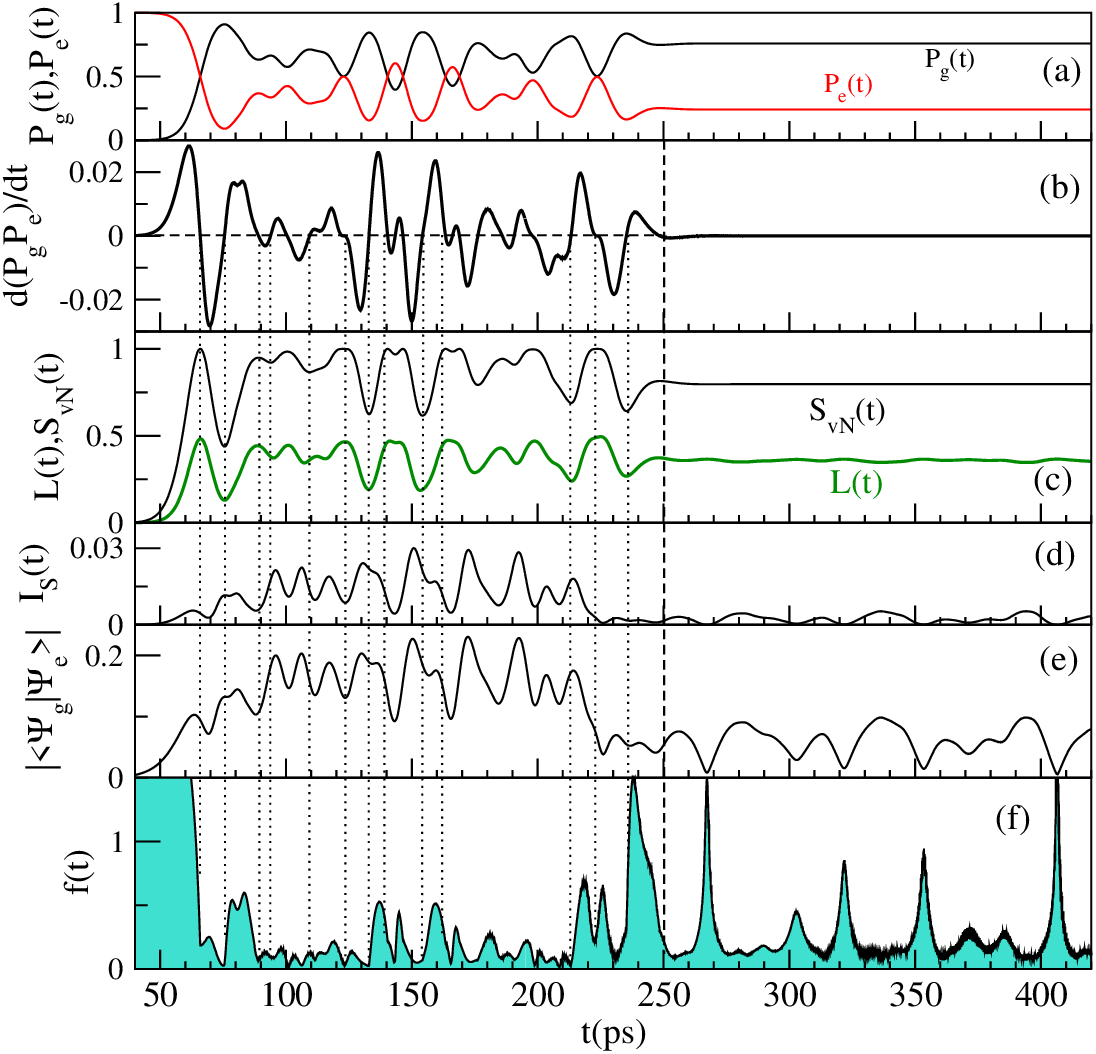}%
 \caption{\label{fig5_4WL} (Color online) Results for a coupling strength $4W_L=13.16$ cm$^{-1}$ between
  the electronic states
 $g=a^3\Sigma_{u}^{+}$ and $e=1_g$ of Cs$_2$ coupled by a pulse with
 the same envelope $e(t)$ shown in Fig.~\ref{fig1_pot}. Evolutions during the pulse and after pulse.
(a) Time evolutions of the
populations $P_g(t)$ and $P_e(t)$. (b) Time evolution of the
 "non-Markovianity factor"  $d( P_g P_e )/dt$.
(c) Time evolutions of the linear entropy $L(t)$ and von Neumann entropy
$S_{vN}(t)$ of the electronic-vibrational entanglement.
(d) Time evolution of the skew information $I_S(t)={\cal I_S}(R,t)/[\Delta V(R)]^2$.
(e) Time evolution of the electronic coherence $C_{l_1}(t)/2=|<\psi_{g}(t)|\psi_{e}(t)>|$.
(f) Non-Markovianity measure $f(t)$. The filled surface shows the integral $\int f(t) dt$.}
\end{figure}

Let us observe more closely the influence exerted by this dynamics of transfer and vibration
on the non-Markovian character of the electronic evolution.
Let us analyze the evolution during the pulse ($t < 250$ ps).
 A first observation (see Figs.~\ref{fig2_WL}(b,c)) is that whenever the electronic-vibrational entanglement 
increases ($dL(t)/dt>0$, $dS_{vN}(t)/dt>0$), the "non-Markovianity factor" is positive, $d( P_g P_e )/dt>0$,
 and whenever entanglement decreases ($dL(t)/dt<0$, $dS_{vN}(t)/dt<0$),
 the "non-Markovianity factor" is negative, $d( P_g P_e )/dt<0$. There is no exception from this rule in this
case, therefore we observe only the situations (2) and (4) from the Table ~\ref{tab:relations}.
Secondly, Figs.~\ref{fig2_WL}(b,c,f) show clearly that, in the time intervals $[t_1,t_2]$ when
the condition of enhanced non-Markovian behavior $d( P_g P_e )/dt>0$ is fulfilled
(i.e. whenever there is
entanglement growth), the total amount of non-Markovianity defined by the integral
 $F(t_1,t_2)$$=\int_{t_1}^{t_2} f(t) dt$ becomes significantly bigger (for example, the
intervals 100-110 ps, 120-130 ps, 145-155 ps, 160-190 ps, or 203-220 ps).
On the contrary, if the entanglement decreases during the time interval $[t_1,t_2]$, $F(t_1,t_2)$ is drastically diminished, 
approaching 0 (between 130-145 ps, for example). 

\begin{figure}
\includegraphics[width=0.95\columnwidth]{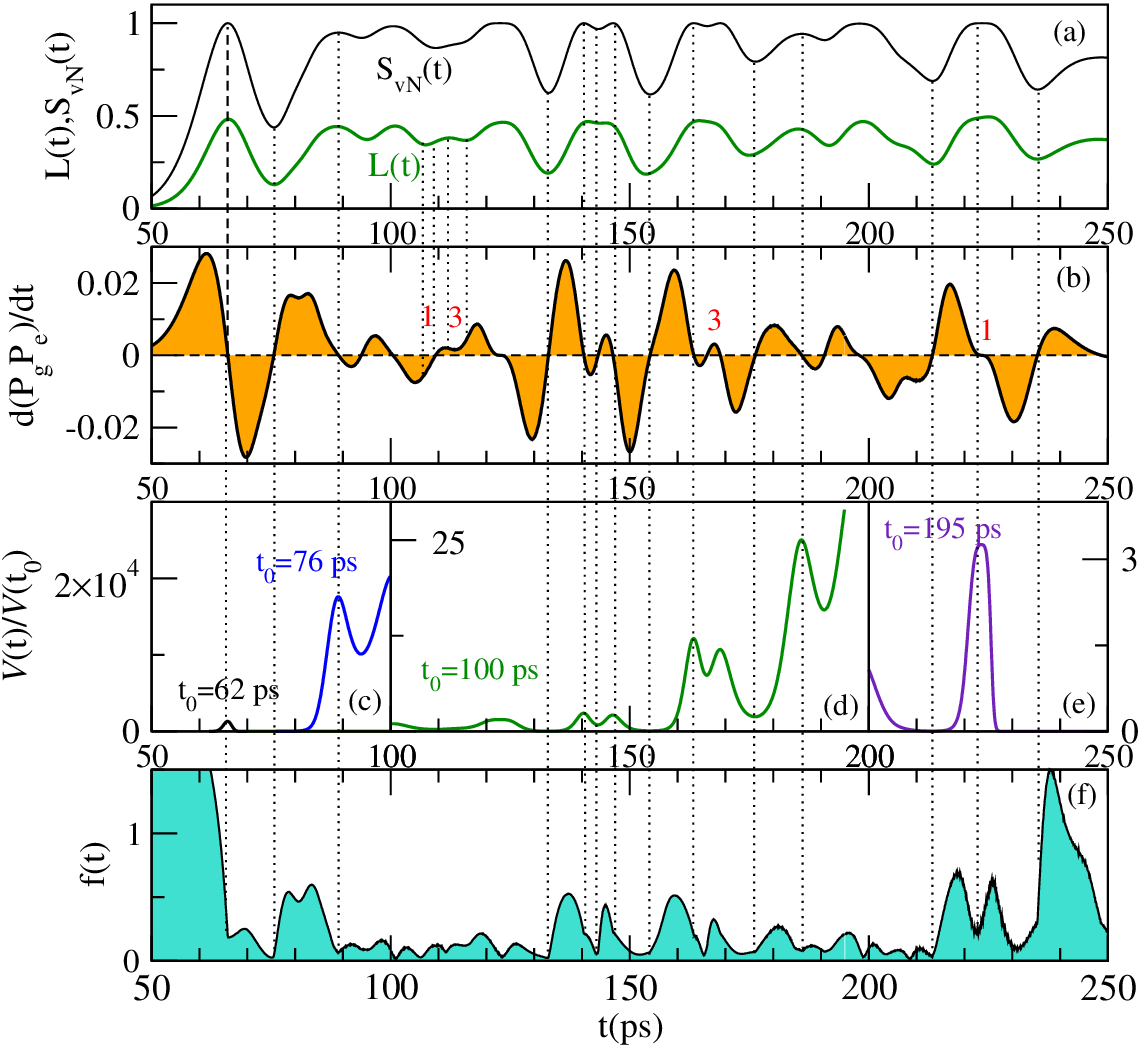}%
 \caption{\label{fig6_nMBlochvol} (Color online) Results during the pulse 
for a coupling $4W_L=13.16$ cm$^{-1}$.(a) Time evolutions of the linear entropy $L(t)$ 
and von Neumann entropy
$S_{vN}(t)$ of the electronic-vibrational entanglement.
(b) Time evolution of the
 "non-Markovianity factor" $d(P_gP_e)/dt$ (non-Markovianity is enhanced for $d( P_g P_e )/dt>0$).
(c-e) Time evolution of the Bloch volume of the accessible states relative to 
the volume at an initial time $t_0$, $\mathcal{V}(t) / \mathcal{V}(t_0)$.
Three time periods (with appropriated initial times $t_0$) are considered: (c)
beginning of the pulse $[50-100]$ ps; (d) the period of constant strength $[100-195]$ ps;
(e) end of the pulse, $[195-250]$ ps.
 (f) Non-Markovianity measure $f(t)$. The filled surface shows the integral $\int f(t) dt$.}
\end{figure}

After pulse ($t > 250$ ps), the electronic populations become constant, and $d( P_g P_e )/dt=0$. 
Vibrational motion in the electronic potentials leads to 
oscillations of the electronic coherence, and implicitly of the linear entropy $L(t)$. 
The non-Markovianity measure is deduced from Eq.~(\ref{gama12cst}) as 
$f(t)=$$\frac{1}{|<\psi_{g}|\psi_{e}>|} \left| \frac{d <\psi_{g}|\psi_{e}> }{dt} \right|$,
taking the form (\ref{ftisolmol}) as function of the electronic coherence $|<\psi_{g}|\psi_{e}>|$.
The results shown in Figs.~\ref{fig2_WL}(e,f) confirm the analysis made in Sec.~\ref{sec:nonmdecohrates}
for a molecule with constant electronic populations: indeed, the non-Markovianity measure
$f(t)$ has minima when the electronic coherence $|<\psi_{g}|\psi_{e}>|$
has maxima (for example, at t=250 ps, 280 ps, 385 ps), and attains maximum values
 when $|<\psi_{g}|\psi_{e}>| \to 0$ (at t=263 ps or 370 ps, for example).
Let us observe the wave packets evolution in Figs.~\ref{fig4_wf2WL}(f-j):
the minima of the electronic coherence are obtained when the overlap of the vibrational wave packets
is minimum. As it can be seen for t=263 ps or 370 ps, the minimum overlap is a result of the
$\psi_{g}(R,t)$ vibration inside the $a^3\Sigma_{u}^{+}$ potential. This vibrational motion
(during which the vibrational wave packets explore the electronic potentials)
diminishes coherence, increasing the electronic-vibrational entanglement and 
bringing a memory character to dynamics. 

\begin{figure}
\includegraphics[width=0.95\columnwidth]{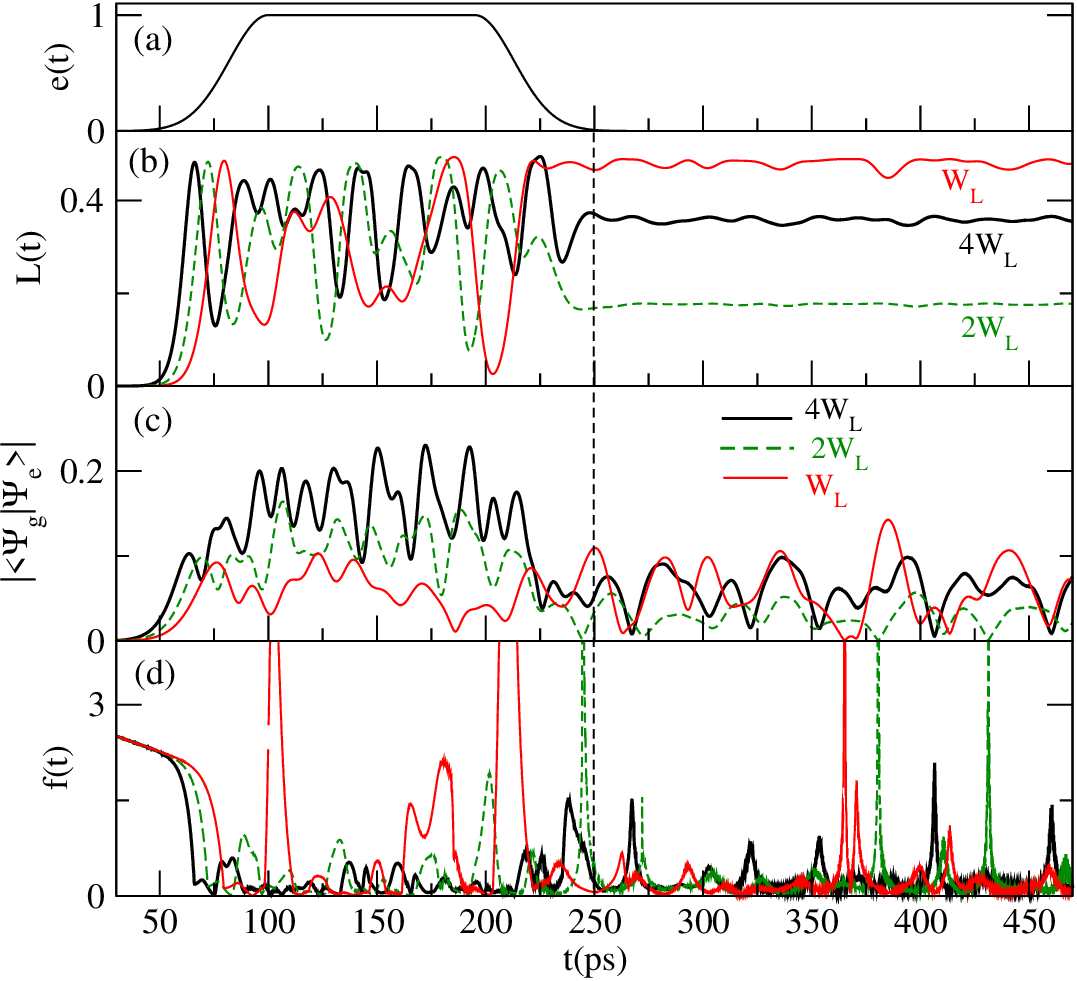}%
 \caption{\label{fig7_3entcohnM} (Color online) Results for the coupling strengths $W_L=3.29$ cm$^{-1}$ (thin line), $2W_L$ (dashed line), and $4W_L$ (thick line) between
  the electronic states
 $g=a^3\Sigma_{u}^{+}$ and $e=1_g$ of Cs$_2$ (Fig.~\ref{fig1_pot}). The dashed vertical line at $t=250$ ps
indicates the end of the pulse.
(a) Pulse envelope $e(t)$. 
(b) Time evolution of the linear entropy $L(t)$ of the electronic-vibrational entanglement. 
(c) Time evolution of the electronic coherence $C_{l_1}(t)/2=|<\psi_{g}(t)|\psi_{e}(t)>|$. 
(d) Non-Markovianity measure $f(t)$.}
\end{figure}

Let us observe the evolution of the two "electronic coherences", $C_{l_1}(t)$ and the skew information
$I_S(t)$, shown in Figs.~\ref{fig2_WL}(e,d), respectively. During the pulse, they
manifest similar behaviors, so we do not observe 
the exceptions signaled in the Table ~\ref{tab:relations} for the cases (2) and (4).
After pulse, their temporal behaviors are also similar,
but $I_S(t) \to 0$ in the time intervals for which $C_{l_1}(t)$ has small values
 (for example, 260-270 ps, or 360-370 ps). At the same time, these intervals are also the periods when
the non-Markovianity measure $F(t_1,t_2)$$=\int_{t_1}^{t_2} f(t) dt$ attains the bigger values after pulse
(see Figs.~\ref{fig2_WL}(d,e,f)).

We will now analyze the results obtained for a much bigger coupling strength, $4W_L=13.16$ cm$^{-1}$, which are shown in
Figs.~\ref{fig5_4WL} (evolution during and after pulse)
and \ref{fig6_nMBlochvol} (detailed evolution during the pulse). The  transfer of population between
 the electronic channels
becomes more intense and fast, and then 
the "non-Markovianity factor"  $d( P_g P_e )/dt$ varies more rapidly (Figs.~\ref{fig5_4WL}(a,b)).
As in the case discussed previously, the increase of the electronic-vibrational entanglement 
($dL(t)/dt>0$, $dS_{vN}(t)/dt>0$) is completely correlated with the positivity of
the "non-Markovianity factor" ($d( P_g P_e )/dt>0$) indicating enhanced non-Markovian behavior.
Also, entanglement decrease corresponds to 
$d( P_g P_e )/dt<0$. The dotted vertical lines in Figs.~\ref{fig5_4WL}(b,c) 
and \ref{fig6_nMBlochvol}(a,b) clearly show these correlations. 
Nevertheless, in this case exceptions from this rule can be observed: indeed, 
as it is shown in Figs.~\ref{fig6_nMBlochvol}(a,b),
one can distinguish small periods of time corresponding to the cases (1) and (3) 
analyzed in the Table~\ref{tab:relations}.
Figs.~\ref{fig6_nMBlochvol}(a,b,f) also show that, as previously, when entanglement increases and 
the condition $d( P_g P_e )/dt>0$ is fulfilled, the integral 
$\int f(t) dt$ is significantly increased.

\begin{table*}
\caption{\label{tab:intft} The total amount of non-Markovianity over the time interval
$[t_i,t_f]$,  $F(t_i,t_f)$$=\int_{t_i}^{t_f} f(t) dt$, calculated for various time intervals 
(during the pulse with the envelope $e(t)$ shown in Fig.~\ref{fig7_3entcohnM}(a), and after pulse),
and for the strengths $W_L=3.29$ cm$^{-1}$, $2W_L$, and $4W_L$ of the coupling.}
\begin{ruledtabular}
\begin{tabular}{cccccc}
&$F$(50,100 ps)&$F$(100,195 ps)&$F$(195,250 ps)&$F$(50,250 ps)&$F$(250,495 ps) \\
\hline
&&&&& \\
$W_L$&57.6&74.6&98.3& 187.3 &53.2 \\
$2W_L$&50.3&20.7&31.6& 102.6  &59.7 \\
$4W_L$&36.8&16.1&22.3&  75.2 &58.9 \\
\end{tabular}
\end{ruledtabular}
\end{table*}

Figs.~\ref{fig6_nMBlochvol}(c-e) show time evolutions of the Bloch volume 
reported at an initial time $t_0$,  $\mathcal{V}(t) / \mathcal{V}(t_0)$,
corresponding to three periods belonging to the time interval $[50,250]$ ps of the pulse action, 
and relative to different initial times $t_0$: (c) beginning of the pulse $[50-100]$ ps
($t_0=62$ ps, $t_0=76$ ps); (d) the period of constant strength $[100-195]$ ps ($t_0=100$ ps);
(e) end of the pulse, $[195-250]$ ps ($t_0=195$ ps).
From the theoretical analysis exposed in Sec.~\ref{sec:nonmdecohrates}, it is expected that
the
Bloch volume will increase, witnessing non-Markovianity, only if $d( P_g P_e )/dt>0$. This is
exactly what we observe in Figs.~\ref{fig6_nMBlochvol}(a-f): increase of the Bloch volume 
is correlated to increase of entanglement, the condition of enhanced non-Markovian behavior $d( P_g P_e )/dt>0$,
and the increase of the integral $\int f(t) dt$.

Non-Markovianity evolution after pulse is shown in Fig.~\ref{fig5_4WL}(f). 
The function  $f(t)$ evolves in the manner previously analyzed, with pronounced
maxima corresponding to the electronic coherence $|<\psi_{g}|\psi_{e}>|$ minima.

The results obtained for three strengths of the coupling ($W_L=3.29$ cm$^{-1}$, $2W_L$, and $4W_L$) 
and the same pulse envelope are compared in Fig.~\ref{fig7_3entcohnM}, which exposes 
the linear entropy $L(t)$ of the electronic-vibrational entanglement, the electronic coherence
$|<\psi_{g}|\psi_{e}>|$, and the non-Markovianity measure $f(t)$. The total amount of non-Markovianity 
$F(t_i,t_f)$$=\int_{t_i}^{t_f} f(t) dt$
over the time interval
$[t_i,t_f]$ was calculated for several time intervals, corresponding
to the beginning of the pulse ($[50,100]$ ps), the period of constant coupling ($[100,195]$ ps),
 the end of the pulse ($[195,250]$ ps), and after pulse ($[250,495]$ ps). The values given in the Table~\ref{tab:intft} show that
the total amount of non-Markovianity corresponding to the pulse action, $F$(50,250 ps), decreases with the increase of the coupling $W_L$, but, after pulse, the values $F$(250,495 ps) calculated for the three strengths of the coupling attain similar values.  

Therefore, we find that during the pulse action, it is 
the weaker pulse which stimulates the bigger amount of  non-Markovianity. This behavior
is related to the Rabi periods of the population exchange between electronic channels, with
a weak coupling enabling a more powerful presence of the vibrational environment. Indeed, a strong coupling  induces a stronger electronic coherence (see Fig.~\ref{fig7_3entcohnM} (c)), favoring the 
transfer of population between channels (localized around the crossing point of the electronic potentials) over the vibrational motion in the molecular potentials. 
A fast transfer of population corresponding to a strong coupling (i.e. small Rabi period) has the effect of "locking" the population in the transfer zone, inhibiting vibration. By contrast, a slower transfer of population, produced by a weak pulse, gives wave packets more time to explore the electronic potentials, increasing gradually the entanglement and
enhancing non-Markovian behavior.

\section{\label{sec:conclu} Conclusions}

We have examined non-Markovian behavior in the reduced time evolution of the electronic subsystem of a laser-driven molecule, as an open quantum system entangled with the vibrational environment. 

Non-Markovianity was characterized using the canonical measures defined in Ref.~\cite{hall2014}
as functions of the negative decoherence rates appearing in the corresponding canonical master equation. 
The canonical measures provide a complete description of non-Markovian behavior, being sensitive to individual decoherence rates when several decoherence channels are present. The Bloch volume of accessible states was also considered as a non-Markovianity witness, even if it does not 
always detect non-Markovian behavior, being only sensitive to the sum of
 the decoherence rates \cite{hall2014}. The use of different non-Markovianity measures helped 
to highlight the enhanced non-Markovian behavior, detected by both measures and generally accompanied by the increase of the electronic-vibrational entanglement.

For a laser-driven molecule
described in a bipartite Hilbert space $\cal{H}$$=$$\cal{H}$$_{el}$$\bigotimes$$\cal{H}$$_{vib}$ with dimension $2 \times N_v$, we have derived the canonical form of the electronic master equation, deducing the canonical decoherence rates  as functions of the electronic populations  $P_g(t), P_e(t)$ and of the electronic coherence (Eqs.~(\ref{gama12}), (\ref{gama3})). Subsequently, the canonical measures of non-Markovianity and the Bloch volume of dynamically accessible states  were obtained.
We found that one of the decoherence rates is always negative, accounting for the inherent non-Markovian character of the electronic evolution.  Moreover, a second
decoherence rate becomes negative if the condition $d( P_g P_e )/dt>0$ is fulfilled, leading to enhanced
non-Markovian behavior, characterized by two negative decoherence rates and a negative sum of the decoherence rates; consequently, the Bloch volume of accessible states increases, detecting enhanced non-Markovian behavior.
Sec.~\ref{sec:nonmdecohrates}  contains a detailed examination of the  canonical measures in relation to the time  evolution of the electronic populations and electronic coherence. 

We showed that in the case of a molecule with constant electronic populations, the non-Markovianity measure $f(t)$ can be seen as a measure of the temporal behavior of the electronic coherence (which determines the evolution of $L(t)$, the linear entropy of entanglement), having minima when the electronic coherence has maxima ($L(t)$ minima), and attaining maximum values whenever the overlap
of the vibrational packets tends to zero ($L(t)$ maxima). This signifies that 
vibrational motion which explore the electronic potentials diminishing nuclear overlap 
(i.e. increasing the linear entropy of entanglement) brings a memory character to dynamics. 

The condition $d( P_g P_e )/dt>0$ was used as an instrument to explore the meaning of enhanced non-Markovian
behavior in the evolution of the electronic subsystem, observing its connections to the dynamics of electronic-vibrational entanglement and electronic coherence in molecule.
We have employed analytical formulas to 
analyze connections between $d(P_g P_e)/dt$, the time behavior of linear entropy of entanglement ($dL/dt$),  and behaviors of speakable and unspeakable \cite{spekkens16} electronic coherences, measured by $l_1$ norm $C_{l_1}(t)$ and skew information ${\cal I_S}(t)$, respectively. 
We have also discussed the possibility of relating the conditions 
$d(P_g P_e)/dt>0$,  $dL/dt>0$, or $d C_{l_1} / dt >0$ to a flow of information from the 
vibrational environment to the electronic open subsystem. In this respect, 
in the appendix we have examined the conditions determining the growth of distinguishability  \cite{breuerlp09} between two electronic states. It appears that the condition $d( P_g P_e )/dt>0$ 
of enhanced non-Markovian behavior participates in the increase of the trace distance 
$D(\hat{\rho}_{el}(t_0),\hat{\rho}_{el}(t))$, and is closely related to the condition of increase of entanglement, $dL(t)/dt>0$.

In the last part of the paper we have analyzed non-Markovian behavior in the reduced evolution
of the electronic states $g=a^3\Sigma_{u}^{+}(6s,6s)$ and $e=1_g(6s,6p_{3/2})$ of the Cs$_2$ molecule,
coupled by a laser pulse. The motion of the vibrational wave packets in the electronic molecular potentials coupled by the laser pulse was simulated numerically, for several strengths of the pulse.
The non-Markovian behavior, characterized using the canonical measures and the Bloch volume, was analyzed in  
relation to dynamics of the electronic-vibrational entanglement and 
 electronic coherence in the molecule. We found that
increase of electronic-vibrational entanglement ($dL(t)/dt>0$, $dS_{vN}(t)/dt>0$) is
correlated with the positivity of
the non-Markovianity factor ($d( P_g P_e )/dt>0$), indicating enhanced non-Markovian behavior, 
with the increase of the Bloch volume, and with the growth of the total amount of non-Markovianity over an interval $[t_1,t_2]$, given by the integral
 $F(t_1,t_2)$$=\int_{t_1}^{t_2} f(t) dt$, where $f(t)$ is the canonical measure of non-Markovianity,
defined from the appearance of negative decoherence rates in the canonical master equation.

We have shown that the total amount of non-Markovianity corresponding to the pulse action decreases with the increase of the coupling. Nevertheless, the values $F(t_1,t_2)$ corresponding to evolutions after pulses 
are similar, probably because 
 analogous domains of vibrational levels are populated, and therefore a similar vibrational dynamics is activated. 
The fact that  during the pulse action, it is 
the weaker pulse which stimulates the bigger amount of  non-Markovianity,
has to be related to the Rabi periods characterizing the exchange of population between electronic channels,
and influencing vibration in the electronic potentials.
 A weak pulse gives more time to vibrational wave packets to explore the electronic potentials, leading to
entanglement increase and enhancement of non-Markovianity.

In conclusion, in a molecule (here with two populated electronic states), the evolution of the electronic subsystem has an {\it inherent non-Markovian character} due to the dynamics of the vibrational environment, even if there is no exchange of population between electronic channels, but only vibrational motion
in the electronic potentials.  {\it Enhanced non-Markovian behavior} of the electronic dynamics 
arises if there is a coupling between electronic channels such that the evolution of electronic 
populations obeys $d( P_g P_e )/dt>0$, and it appears as a dynamical property 
associated with the {\it increase of the electronic-vibrational entanglement}.
Several non-Markovianity regimes, determined by the sign of the non-Markovianity factor $d( P_g P_e )/dt$, were analyzed in Sec.~\ref{sec:nonmdecohrates} and Sec.~\ref{sec:nMarkov-entcoh}. 

A key motivation shaping the present work was to examine
non-Markovian behavior of the electronic evolution in relation to the
dynamics of the quantum correlations in the molecular system. In this sense,
observation of the correlation phenomena accompanying  enhancement of non-Markovianity reveals 
appropriate ways to understand non-Markovian behavior.
Therefore, if the non-Markovian character of the electronic dynamics cannot be separated from the presence of the electronic coherence,
the most significant relation is between non-Markovianity and entanglement dynamics: 
We have shown that non-Markovianity of the electronic evolution is essentially a dynamical property generated during the increase of electronic-vibrational entanglement.

\begin{acknowledgments}
This work was supported by the LAPLAS 4 and LAPLAS 5 programs of the Romanian National Authority
for Scientific Research.
\end{acknowledgments}

\appendix*

\section{\label{sec:appendix} Distinguishability between two electronic states, $\hat{\rho}_{el}(t_0)$
and $\hat{\rho}_{el}(t)$}

Distinguishability between two electronic states $\hat{\rho}_{el}(t_0)$
and $\hat{\rho}_{el}(t)$ can be analyzed using as measure the trace distance $D(\hat{\rho}_{el}(t_0),\hat{\rho}_{el}(t))$
between the two states, defined as \cite{breuerlp09,breuer12}
\begin{equation}
D(\hat{\rho}_{el}(t_0),\hat{\rho}_{el}(t)) = \frac{1}{2} \text{Tr}_{el} |\hat{\rho}_{el}(t_0)-\hat{\rho}_{el}(t))|.
\label{trdistance}
\end{equation}
Taking into account the matrix of the electronic density given by Eq.~(\ref{matdensel2}),
one obtains \cite{breuer12}
\begin{equation}
D(\hat{\rho}_{el}(t_0),\hat{\rho}_{el}(t)) = \sqrt{  [P_g(t_0)  - P_g(t)]^2 + | C (t_0)-C (t)  |^2 }.
\label{distel0t}
\end{equation}
In Eq.~(\ref{distel0t}), $P_g(t_0)  - P_g(t)$ is the difference of the populations between $t_0$ and $t$, and 
$C (t_0)-C (t)$ is the difference between the complex nondiagonal elements 
$C (t)$$=<\psi_{g}(t)|\psi_{e}(t)>=|C(t)|$exp$[i \alpha(t)]$ of the electronic density matrix (\ref{matdensel2}) 
at $t_0$ and $t$. The $l_1$ norm measure of the electronic coherence is $C_{l_1}( \hat{\rho}_{el})$$=2|C(t)|$ . 

We look for the conditions determining an increase of the trace distance, i.e. a positive rate of change
$dD(\hat{\rho}_{el}(t_0),\hat{\rho}_{el}(t))/ dt >0$. From Eq.~(\ref{distel0t}) one obtains
the following equation giving the rate of change of the trace distance, $dD(\hat{\rho}_{el}(t_0),\hat{\rho}_{el}(t))/ dt$:
\begin{eqnarray}
D(\hat{\rho}_{el}(t_0),\hat{\rho}_{el}(t)) \frac{d D(\hat{\rho}_{el}(t_0),\hat{\rho}_{el}(t)) }{dt}  \nonumber \\
=     [P_g(t)  - P_g(t_0)] \frac{dP_g(t)}{dt} + |C(t)| \frac{d |C(t)|}{dt}  \nonumber \\
- |C(t_0)| \frac{d |C(t)|}{dt} \text{cos} [\alpha(t_0) - \alpha(t)] \nonumber \\
- |C(t_0)||C(t)| \text{sin} [\alpha(t_0) - \alpha(t)] \frac{d \alpha(t)}{dt}.
\label{dtdistel}
\end{eqnarray}

As it could be expected, Eq.~(\ref{dtdistel}) shows that $d D(\hat{\rho}_{el}(t_0),\hat{\rho}_{el}(t)) / dt$ is an oscillating function, which becomes positive or negative depending on the evolution at the instant $t$ and on the
initial state at $t_0$.  Nevertheless, some interesting observations can be made. 

Let us consider the right hand side of Eq.~(\ref{dtdistel}).
The first term becomes positive, $[P_g(t)  - P_g(t_0)] dP_g(t)/dt >0$, if 
$\text{sgn}(dP_g / dt)$$=\text{sgn}[P_g(t)-P_g(t_0)]$, i.e.  on those intervals $[t_0,t]$ of the time evolution
on which a smaller population  at $t_0$ is increased at $t$ ($P_g(t_0) < P_g(t)$, $dP_g(t)/dt >0$)
or a larger population  at $t_0$ is diminished at $t$ ($P_g(t_0) > P_g(t)$, $dP_g(t)/dt <0$). 
In Sec. \ref{sec:nonmdecohrates} we have shown that  the condition $(P_g-P_e)$$dP_g/dt$ $<0$ 
of enhanced non-Markovian behavior is fulfilled when the transfer of population
 between the two electronic channels is such as the larger population decreases
 (i.e. the smaller electronic population increases). Moreover, this is also the condition leading to
the increase of the electronic-vibrational entanglement. Therefore, our observation is that on time intervals 
$[t_0,t]$ when the condition  $(P_g-P_e)$$dP_g/dt$ $<0$  ($d( P_g P_e )/dt>0$) is fulfilled,
 also $[P_g(t)  - P_g(t_0)] dP_g(t)/dt >0$. 

The second term on the right hand side of Eq.~(\ref{dtdistel}) is equal to $(C_{l_1}d C_{l_1} / dt)/4$,
and it becomes positive if the electronic coherence increases, $d C_{l_1} / dt >0$.

The last two terms 
on the right hand side of Eq.~(\ref{dtdistel}) depend on the "complex coherences" $C (t_0)$ and $C (t)$, 
and can be characterized as "easily oscillating" terms, whose signs are rapidly changing. 

Let us suppose that the electronic state $\hat{\rho}_{el}(t_0)$ is a state with electronic coherence $|C(t_0)|=0$.
Therefore, the last two terms become 0, and Eq.~(\ref{dtdistel})
 shows that the trace distance between $\hat{\rho}_{el}(t_0)$ and another state
$\hat{\rho}_{el}(t)$ will increase
 ($dD(\hat{\rho}_{el}(t_0),\hat{\rho}_{el}(t))/ dt >0$) in a interval
$[t_0,t]$ in which the conditions $d( P_g P_e )/dt>0$ and $d C_{l_1} / dt >0$
are fulfilled. In other words, distinguishability between $\hat{\rho}_{el}(t)$ and a state 
$\hat{\rho}_{el}(t_0)$ with coherence $C_{l_1}(t_0)=0$ is increased when $d( P_g P_e )/dt>0$ and $d C_{l_1} / dt >0$.
Breuer, Laine and Piilo \cite{breuerlp09} interpret the growth of distinguishability between two states
of the open system as the signature of a reversed flow of information from the environment back to the open system,
an essential trait of non-Markovian behavior.

\bibliography{art-nonmarkov}

\end{document}